\documentclass[review]{elsarticle}
\usepackage{booktabs}
\usepackage{amsmath}
\usepackage{mathrsfs}
\usepackage[ruled]{algorithm2e}
 \usepackage{graphicx,rotating,booktabs,multirow, array}
\usepackage{csquotes}
\usepackage[table]{xcolor}
\usepackage{siunitx}
\usepackage{pgfplots}
\usetikzlibrary{positioning}
\usepackage{lscape}
\usepackage{setspace}
\usepackage{etoolbox}
\usepackage{booktabs}
\usepackage{makecell}  
\usepackage{tabularx}
\usepackage{tikz}
\usetikzlibrary{shapes}
\usetikzlibrary{plotmarks}

\makeatletter
\def\ps@pprintTitle{%
  \let\@oddhead\@empty
  \let\@evenhead\@empty
  \def\@oddfoot{\reset@font\hfil\thepage\hfil}
  \let\@evenfoot\@oddfoot
}
\makeatother

\usepackage{extarrows}
 \usepackage[a4paper, total={6in, 8in}]{geometry}

\usepackage{algpseudocode}

\usepackage{hyperref}
\usepackage{epstopdf}
\usepackage{textcomp}
\AppendGraphicsExtensions{.tif}

\begin{document}
\begin{frontmatter}

\title{Wall shear stress and pressure patterns in aortic stenosis patients with and without aortic dilation captured by high-performance image-based computational fluid dynamics }

\author[mymainaddress]{Hadi Zolfaghari\corref{cor1}}
\ead{hz382@damtp.cam.ac.uk}
\cortext[cor1]{Corresponding Author}


\author[my3rdmainaddress]{Mervyn Andiapen}
\author[my2ndmainaddress,my3rdmainaddress]{Andreas Baumbach}
\author[my2ndmainaddress,my3rdmainaddress]{Anthony Mathur}
\author[mymainaddress]{Rich R.  Kerswell}

\address[mymainaddress]{Department of Applied Mathematics and Theoretical Physics, University of Cambridge,  Cambridge, UK }
\address[my2ndmainaddress]{Barts Heart Centre, Barts Health NHS Trust, London, UK
}
\address[my3rdmainaddress]{Centre for Cardiovascular Medicine and Devices, William Harvey Research Institute, Queen Mary University of London, London, UK}

\begin{abstract}

 Spatial patterns of elevated wall shear stress and pressure due to blood flow past aortic stenosis (AS) are studied using GPU-accelerated patient-specific computational fluid dynamics.  Three cases of moderate AS,  one with a dilated ascending aorta and two within the normal range (root diameter less than 4cm)  are simulated for physiological beat cycle waveforms obtained from echocardiography data. The computational framework is built based on sharp-interface Immersed Boundary Method,  where aortic geometries segmented from CT angiograms are integrated into a high-order incompressible  Navier--Stokes solver.  We show that even though the wall shear stress is elevated and oscillatory due to turbulence in the ascending aorta for all the cases,  its spatial distribution is significantly more focused for the case with dilation than those without dilation.  This focal area is linked to aortic valve jet impingement on the outer curvature of the ascending aorta,  and has been shown $in~vivo$ using 4D flow MRI of aortic stenosis patients with aortic dilation (\textrm{\textit{van Ooij et al., J. Am. Heart Assoc., 6(9), 2017}}).  We show that this focal area also accommodates a persistent pocket of high pressure, which is likely to have contributed to the dilation process through an increased wall-normal forcing.  The cases without dilation,  on the contrary, showed a rather oscillatory pressure behaviour,  with no persistent pressure \enquote{buildup} effect.  We further argue that a more proximal branching of the aortic arch could explain the lack of a focal area of elevated wall shear stress and pressure, because it interferes with the impingement process due to fluid suction effects.  These phenomena are further illustrated using an idealized aortic geometry.  We finally show that a restored inflow condition eliminates the focal area of elevated wall shear stress and pressure zone from the ascending aorta.

\end{abstract}

\begin{keyword}
Cardiovascular flow, image-based simulations, patient-specific modelling, turbulent flows, hybrid supercomputing
\end{keyword}

\end{frontmatter}

\section{Author summary}
Dilation of the aorta can be a precursor to fatal events such as aortic dissection.  Aortic dilation is an involved process which depends on the biological properties of the aorta as well as on the aortic blood flow features. Recent flow imaging studies suggest that a stenotic (i.e., narrowed) aortic valve leads to a turbulent flow character which elevates the fluid stresses applied onto the aortic wall. These excessive stresses are then linked to dilation of the aorta through weakening of the vessel wall tissue. Not all patients with stenotic aortic valves develop aortic dilation, however. Aortic geometry may be a key factor in whether or not a stenotic valve leads to dilation, as it affects the way the restricted aortic flow jet interacts with the vessel wall. We investigate the role of this factor by studying three aortic stenosis patients with and without aortic dilation. To this end, we set up and use a high-fidelity image-based computational framework for detailed simulation of the turbulent flow states and show that certain geometrical layouts of the aorta may be more likely to lead to dilation, because they create persisting and focused areas of excessive blood flow stress on the aortic wall.

\section{Introduction} 


\subsection{Background: chaotic character of the aortic stenosis flow}

Blood flow through a stenotic or narrowed aortic valve is characterized by a pulsatile and moderately high velocity jet inside a relatively small and curved tube.  The resulting flow is unsteady, irregular and chaotic \cite{stein1976turbulent} which comprises a wide range of time and length scales. Therefore, the aortic stenosis (AS) flow can be said to be turbulent \cite{pope2000turbulent}.  This turbulent character impacts the aortic wall as well as the blood components downstream of the valve as we lay out next.  It also acts to enhance itself through increasing the thrombosis potential around the narrowed valve,  which results in an even more severe stenosis, and thereby stronger turbulence. 

Turbulence has been shown to deteriorate the vascular wall and thereby trigger aortic dilation or aneurysms.  Davies et al. \cite{davies1986turbulent} by means of an \textit{in vitro} study showed that turbulent shear stresses as low as 1.5 $\si{dyne/cm^2}$ can trigger substantial endothelial cell DNA release, misalignment and loss, while laminar shear stresses of magnitude 8.5 $\si{dyne/cm^2}$ did not initiate any cell cycle even at much longer exposure times.  Other works have highlighted the role of an oscillatory or turbulent wall shear stress character in increasing the risk of atherosclerotic lesions \cite{traub1999shear, lehoux2003cellular}.  The elevated wall shear stress in a turbulent flow environment has since been used as an important risk stratification parameter in numerous studies involving an impaired aortic flow \cite{van2017aortic, pasta2017silico,  jr2011computational, kauhanen2019aortic} or other diseased arterial flow scenarios such as for intracranial aneurysms and stenoses \cite{valen2018real, morbiducci2020wall,  szajer2018comparison, zhang2018study}.  Of particular importance is the large $in~vivo$ 4D flow MRI study of van Ooij et al.  \cite{van2017aortic}, where a focal zone of high wall shear stress on the distal outer curvature of ascending aorta (denoted by AAo hereinafter) was shown to be a hallmark of AS in patients with present aortic dilation, which over-rides the role of valve phenotype (i.e.,  whether the diseased valve was bicuspid or tricuspid).  They further argued that the apparent AAo dilation could be due to this \textit{focal} area, which has been reported in several studies of the bicupid valve hemodynamics,  where the dilation of AAo is more common than the aortic stenosis patients with a tricuspid valve \cite{wilton2006post}.  This zone, although more extended towards the proximal AAo,  has also been captured in recent numerical simulations of aortic valve stenosis \cite{manchester2021analysis},  which addressed a case with present dilation.  The hemodynamic evidence on aortic stenosis cases without dilation is more scarce.  Understanding the hemodynamics differences between aortic stenosis cases with and without dilation can help quantify the role of flow next to other relevant factors (such as mechanobiology of the arterial wall tissue) in terms of future risk of developing aortic dilation.

\subsection{Scope of this work}

The present study resorts to high-order and high resolution image-based CFD simulations to investigate the effect of aortic geometry on the hemodynamics features of the aortic stenosis flow.  Three patients  who were candidates for transcatheter aortic valve implantation (TAVI) are studied.  One patient with ascending aortic dilation is compared against two cases of minimal dilation. Dilation is defined based on the aortic root diameter, where root diameters above 4cm are considered to be dilated \cite{zhu2020surgical} (note that criticality of dilation is not addressed here). For this present study, we particularly note the focal area of high wall shear stress on the outer surface of the ascending aorta for the mild to severe AS patients as shown in the large 4D flow MRI study of Van Ooij et al.  \cite{van2017aortic}.  This zone which was linked to dilation of the ascending aorta (aortopathy),  has been shown in several studies concerning aortic stenosis of the bicuspid valves (see \textit{in vivo} study of  \cite{garcia2019role} and \textit{in silico} study of \cite{youssefi2017patient} as examples).  In the case of tricuspid valve with stenosis,  this, to our knowledge,  has been shown only in CFD study of \cite{manchester2021analysis} (who studied a patient with tricuspid valve stenosis and aortic dilation).  Little has been reported about hemodynamics of tricuspid aortic valve stenosis patients who do not develop aortic dilation.  In particular, it is unclear why the elevated wall shear stress due to a turbulent flow character in the AAo for these cases does not lead to aortic dilation.  It is therefore useful to investigate the hemodynamics for these patients next to cases with aortic dilation,  to understand the underlying flow-mediated factors behind the dilation.

We first present an image-based multi-GPU-accelerated blood flow simulation tool based on the incompressible Navier-Stokes equations, aiming at high resolution simulations of chaotic flow in patient-specific aortic geometries.   The GPU-accelerated incompressible Navier--Stokes solver with immersed boundary method was presented previously in \cite{zolfaghari2021high}.  Here,  the solver is upgraded with an image-based module,  which works out complex arterial geometries obtained from CT or MRI scans.  The code is also refurbished for integration of velocity and pressure waveform data within the computational domain.  As a result,  the presented simulation code takes 3D texture model of the aortic geometries extracted from CT images, together with velocity and pressure waveforms (obtained from echocardiography and Windkessel models respectively) as inputs, and returns detailed flow field outputs including, velocity, pressure, and Lagrangian coherent structures.  Second-order quantities such as wall shear stress are supported using the computational geometry toolkits and the Python interface of the open source software Paraview (\href{https://www.paraview.org}{www.paraview.org}).  The simulation data are resolved both in space (the grid resolution matches the CT scan resolution, or goes beyond it depending on flow characteristics) and time (numerous flow outputs, targeting a maximum frequency of 1400 Hz are generated which amount to roughly 7TBs of data per simulation of a single heart-beat). The sharp-interface Immersed Boundary technique is used for integration of the complex geometry into a Cartesian simulation domain \cite{mittal2008versatile,mittal2005immersed, zolfaghari2017simulations}. One key advantage of this current implementation compared to more classical Immersed Boundary implementations is that, no surface mesh for the solid is required here. Instead the computational geometric features such as unit normals and boundary cells are directly obtained from a level-set-like map that is obtained from the segmentation output. This offers a superiority in terms of computational performance (no load balancing issues stemming from a local Lagrangian mesh, as the level-set map is defined globally and decomposed equally across parallel processes) and geometrical pre-processing (it no longer requires a meshing step after a computer-aided design file is produced from the segmentation step). Although the current implementation only deals with rigid wall models, it can be further extended to include moving walls inferred from MRI images in a direction prescribed motion fashion  \cite{jin2003effects}, or through an extension to fluid-structure interactions \cite{lin2022eulerian,rycroft2020reference}. The rigid body assumption which is used in various studies in the literature \cite{manchester2021analysis,lantz2013numerical,cheng2010analysis, numata2016blood,assemat2014three,prahl2016effects},  provides a good approximation for a wide range of patients with valvular and aortic diseases, as this cohort usually corresponds to a large degree of atherosclerotic plaques is associated with higher stiffness of the aortic wall \cite{van2001association, cecelja2012role}. Further, as illustrated in \cite{reymond2013physiological} aortic flow models based on rigid walls provide a good approximation of wall shear stress when the wall deformation is below 10-15\%.

Our simulations reveal a focal area of elevated wall shear stress on the outer curvature of the AAo for the AS patient with aortic dilation, which is in agreement with findings of \cite{van2017aortic}.  However,  for the cases without dilation, even though the wall shear stress was elevated in the AAo,  it was distributed more evenly in the circumferential as well as streamwise directions.  In line with the hypothesis that this area develops due to aortic valve jet impingement on the aortic wall for the dilated case,  we showed that a temporally persistent and focal zone of high pressure accompanies the elevated shear stress area.  On the other hand, the cases with root diameters within the normal range showed no persistent high pressure zones.  This focal area appeared to be spatially more oscillatory with respect to the wall shear stress values, but less so with regard to pressure.  It could therefore be hypothesized that this coherent high pressure zone might as well contribute to the dilation of the vessel by supplying the normal force for the deformation,  while high wall shear stress and its oscillatory nature,  weaken the aortic wall matrix to ease such deformation.  The difference in spatial distribution of wall shear stress and pressure patterns is then investigated through volume rendering of the instantaneous flow fields.  It became apparent that the dilated aorta allows for aortic valve jet impingement on the aortic wall which results in a turbulent zone with elevated shear stresses, and a high pressure zone due to adjacent flow deceleration.  However,  the aortic jet impingement location for the cases with no dilation seems to coincide with the brachiocephalic artery inlet.  The suction effect by this artery further disrupts the otherwise focal jet impingement process,  leading to more severe turbulent flow with a farther extended impact on the aortic wall.  This stronger turbulence character is verified using pressure signals taken within the AAo.  We tested this latter hypothesis using a mock model,  where we created a simplified aortic geometry with a more distal brachiocephalic artery inlet, and showed that the focal zone of high wall shear stress and pressure was present due to undisturbed aortic jet impingement.   Further case studies and comparison to \textit{in vivo} data should be conducted to validate this observation.


  Finally,  we show that a healthy inflow condition (i.e.,  with no AS) leads to no focal zone of elevated WSS and pressure in the dilated case.  To this end,  we simulated a \enquote{restored} flow scenario, where the aortic inflow (i.e., after aortic valve replacement) was modelled by tripling the aortic jet area while maintaining the peak flow rate.  It is shown that a restored flow eliminates the focal area of excessive wall shear stress and pressure from the AAo.  This result is verified even under extreme stress, where we doubled the peak aortic valve jet velocity.  

\section{Clinical data acquisition for the study group}
Three patients with aortic valve stenosis were recruited from St Bartholomew’s Hospital (London, UK) and CT scan were performed on a third-generation dual-source system (SOMATOM Force, Siemens, Forchheim, Germany). The study received ethical approval from the Health Research Authority and Westminster Research Ethics Committee (18/LO/1583) and was sponsored by the Queen Mary University of London with Joint Research and management Office, as defined under the sponsorship requirements of the UK Policy framework for Health and Social Care Research (2017) and ICH GCP.

\section{Simulation procedure}
\label{secFlowSolver}
\subsection{Governing equations}


We solve the dimensionless incompressible Navier--Stokes equations

\begin{equation} 
\label{eq1}
\frac{\partial{}}{\partial{t}}\begin{bmatrix}
    \textbf{\underline{u}} \\
    0 
\end{bmatrix}+
\begin{bmatrix}
    \mathscr{-L}       & \mathscr{G} \\
    \mathscr{D}      & 0
\end{bmatrix}
\begin{bmatrix}
    \textbf{{\underline{u}}} \\
    p 
\end{bmatrix}=
\begin{bmatrix}
    \mathscr{N} \textbf{{\underline{u}}} \\
    0 
\end{bmatrix},
\end{equation}


\noindent where $\textbf{{\underline{u}}}$=[$u_x$,$u_y$,$u_z$] and \textit{p} denote dimensionless velocity and pressure, respectively. Operators $\mathscr{G}$, $\mathscr{D}$, $\mathscr{L}$, and $\mathscr{N}$, represent the differential operators $\nabla$, $\nabla \cdot$, $Re^{-1}{\nabla}^2$ and $-{\textbf{\underline{u}}} \cdot {\nabla}$, respectively. The Reynolds number ($Re$) is defined as:
 
\begin{equation} 
\label{eq2}
Re = \frac{{\mathscr{U}_0\mathscr{L}_0}}{\nu},
\end{equation}

\noindent where $\mathscr{U}_0$, $\mathscr{L}_0$ and $\nu$ are the reference velocity, length scale and blood kinematic viscosity.  The flow equations \eqref{eq1} are discretized using explicit 6th-order finite-differences in space and a low-storage 3rd-order Runge-Kutta scheme \cite{wray1986very} in time. The solver has been optimized for task parallelism on multicore distributed memory supercomputers \cite{henniger2010high} as well as for combined data- and task-parallelism on multicore-manycore GPU-based hybrid-node supercomputers \cite{zolfaghari2019high,zolfaghari2021high}. It has been utilized for a range of challenging laminar-turbulent transition flow scenarios \cite{henniger2010high,john2014stabilisation, john2016secondary, zolfaghari2019absolute,  zolfaghari2022sensitivity}. The GPU-accelerated version of the solver which is up to two orders of magnitude faster than the CPU-based paralllel solver,  is particularly favourable for ultimate usage in a clinical setting,  as it reduces a single-beat simulation time from months to 1-3 days at the resolution levels used in this work,  using only 8 GPUs.

\subsection{Computer model of the aorta: geometry and boundary conditions}

\subsubsection{Aortic geometry}

\begin{figure}
\centering
\includegraphics[width=0.9\textwidth]{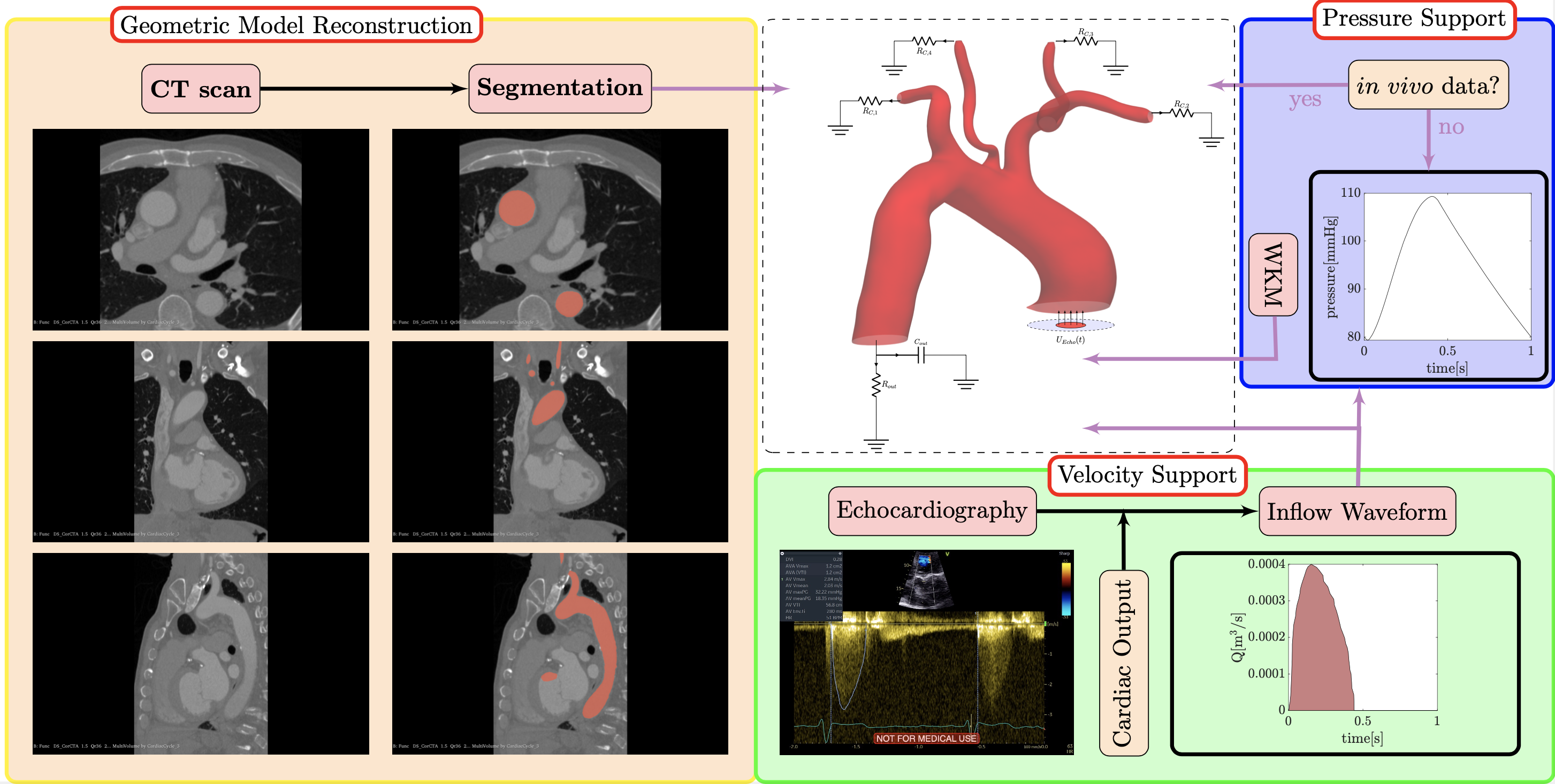}
\caption{Workflow of creating the digital input model (shown inside the dashed panel) for the patient-specific blood flow simulations.  Geometric Model Reconstruction: this block (on the left) shows how CT angiogram data (in DICOM format) are segmented in 3D slicer software (where the region of interest is marked up) and then converted in to a 3D volumetric model of the aorta.  Velocity Support:  the velocity waveform is extracted from echocardiography data and is used directly at aortic inflow boundary.  It is then converted, using a reference cardiac output,  to a flow-rate waveform which is required for construction of a pressure waveform.  Pressure Support: the numerical solver is capable of assimilation of $in~vivo$ pressure data, if available,  into the solution (this is done rigorously by modifying the Poisson equation for pressure).  In absence of this data,  a surrogate model of the outflow pressure is constructed based on the Windkessel model (WKM).  In this study,  a pure velocity waveform is used at the inlet,  a pure pressure waveform (based a two-element WKM) is used at the descending aorta outlet,  and adaptive zero-stress boundary conditions are applied in the neck artery outflows.}
\label{fig:fig1}
\end{figure}

A geometrical model of the aorta is generated through two steps: i) medical image processing, mainly the segmentation of CT angiography scans of the patient's chests and ii) creation of the 3D texture file , i.e. converting the segmentation result into a geometrical format,  such as a visualization toolkit vtk (\href{https://vtk.org}{www.vtk.org}) or stereolithography (stl) file,  which is suitable for integration into the flow solver (see Figure \ref{fig:fig1} under "geometric model reconstruction").  CT angiograms of three patients with severe aortic stenosis who were selected for transcatheter aortic valve implantation (TAVI) in Barts Heart Centre (Barts Health NHS Thrust) are used in this study.  CT datasets in DICOM format are then processed in 3D Slicer software (\href{https://slicer.org}{www.slicer.org}), where a region of interest (ROI) including the thoracic aorta is segmented to yield an isolated geometrical model of the aorta.  Second, the segmentation outputs are reorganized as a standard geometry file which are suitable for integration in a Cartesian grid Immersed Boundary flow solver. The spatial resolution of the data could be adjusted via a standard upsampling or downsampling procedure. A summary of the patient CT data and geometrical model outputs is given in Table \ref{tab:dimsCT}.

\begin{figure}
\centering
\includegraphics[width=0.7\textwidth]{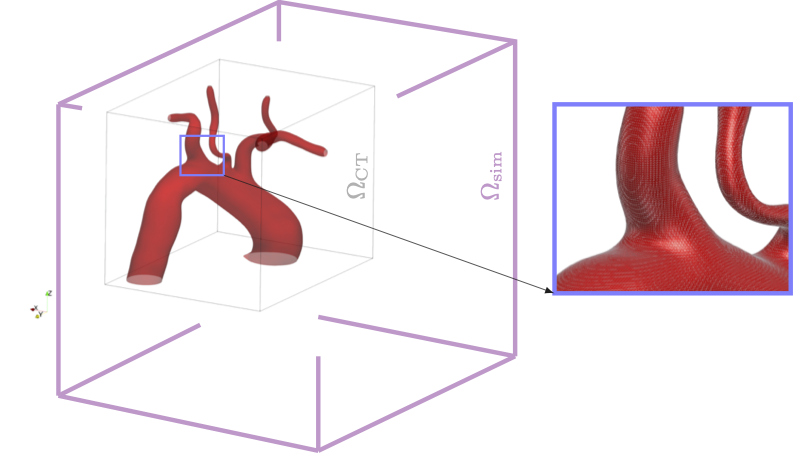}
\caption{Simulation domain $\Omega_\text{sim}$ and the embedded subset of interest $\Omega_\text{CT}$.  The $\Omega_\text{CT}$ includes the segmented model geometry,  all velocity and pressure support data are applied over fringe zones outside of $\Omega_\text{CT}$, as volumetric patches attached to the inlet and outlets of the model.  The flow equations are then solved for the whole $\Omega_\text{sim}$.  The blue window which is magnified on to the right side of the simulation domain shows the spatial resolution represented by thin mesh lines on the surface.  }
\label{fig:fig2}
\end{figure}

\subsubsection{Inflow boundary conditions based on spectral echocardiography data}

For each patient, a cardiac waveform was extracted from the Continuous Wave (CW) spectral echocardiography data (see Figure \ref{fig:fig1} under "Velocity Support"),  and a heart rate of 60 beats per minute (bpm) is considered. The blood flow velocity past the aortic valve (AV) is then converted to a discrete waveform in CSV format to be read as an input for the numerical simulations. 

The aortic jet issuing from the healthy or diseased aortic valve may take complex geometrical shapes depending on a variety of factors including the anatomy of the aortic root and the aortic valve. For calcified valves this becomes even more complex, where one or more valve cusps might become partially or fully immobile which leads to irregular cross-sectional jet profiles. This leads to a diverse set of incoming jet profiles, ranging from almost circular shapes to triangular, hexagonal or three-point-star profiles, as well as more eccentric combinations of these in presence of calcification. 4D flow MRI imaging of the flow profiles behind an aortic valve with stenosis indicates a peak flow velocity region roughly at the centre of the vessel's cross section \cite{manchester2021analysis, archer2020validation}. It also shows that the forward flow part of the profile sustains its shape while accelerating and then decelerating in amplitude during the systolic phase, and loses its shape only when the valve is closed. Given these observations, and because the majority of turbulent flow events occur during the systolic phase, we define the inflow profile $\Omega_{in}(x,y)$ to include two parts: the core circular area $\Omega_{jet}$ with radius $r_{jet}$ indicates the jet issuing from the valve and an outer part $\Omega_{cusp}$ which fills the region between the $\Omega_{jet}$ and the aortic lumen ($\Omega_{in} = \Omega_{jet} \cup \Omega_{cusp}$).  Note that here the inflow cross-section is a subset of $x-y$ plane, as the $z-$ axis has been locally aligned with the vessel walls at the inlet. We assume the flow profile at the inlet to be uni-directional, therefore the inflow velocity field at the inlet is purely in the $z-$direction. A plug velocity profile is then imposed in $\Omega_{in}$ with the $z$-component $w$ of

\begin{equation}
\centering
w(x,y,t) =
\begin{cases}
    U_{echo}(t) & \text{if } x\in \Omega_{jet},\\
    0           &   \text{if}~ x\in \Omega_{cusp}.
\end{cases}
\end{equation}
 
\noindent The radius of the core area is obtained based on a standard peak flow rate of $Q_{peak}=0.0004\si{{m}^3/s}$ for all cases in the paper. With the peak velocity $U_{peak}$ obtained from the echocardiography data, the radius of the aortic jet is defined as 

\begin{equation}
\centering
r_{jet} = \sqrt{\frac{Q_{peak}}{4\pi U_{peak}}}= {\frac{1}{100\sqrt{\pi U_{peak}}}}.
\end{equation} 

\noindent Finally, note that peripheral part ($\Omega_{cusp}$) of the inlet is not annular, because the lumen shape is not circular. Therefore, we have to also define a center for $\Omega_{in}$ to complete our definition. We define this point $\mathbf{C} = (x_C, y_C)$ based on the first moment of area of $\Omega_{in}$,  that is

\begin{equation}
\centering
\mathbf{C} = (x_C, y_C) = \frac{1}{\Omega_{in}}\int_{\Omega_{in}}(x,y)~d\Omega.
\end{equation}

The inflow profile model described above has been fabricated based on a series of assumptions, because the phase-contrast flow MRI information at the inflow was not present. Such date can help achieve a better model of the aortic inflow profile, and will be a subject for our future work.

%
%
\subsubsection{Outflow boundary conditions}

Boundary conditions (BC) applied on the outflow boundaries include Dirichlet-type homogeneous boundary conditions at the descending aorta outflow boundary and zero-stress compatibility-type pressure boundary conditions at the outflow boundaries of the neck arteries. The Dirichlet type BC values for pressure at the descending aorta outflow are calculated dynamically based on a two-element Windkessel model (2EWKM) \cite{westerhof2009arterial}. These values construct a waveform which is obtained from a flow rate waveform input which has a peak value that is a fraction of that for the inflow waveform (a standard value for this fraction is 0.85 which assumes that roughly 5\% of the aortic flow is supplied to each of the neck arteries \cite{tse2013computational}). Note that the these fractions have only been used to obtain the outflow pressure waveform for the descending aorta. The zero stress BCs that are applied to the neck arteries merely ensure that a sum of flow fraction supplied to these arteries amounts to 15\% of the incoming flow (while the flow fraction to each artery can be different than 5\%).  Based on this flow waveform, a pressure waveform is generated by time-integration of the flow rate-pressure equation arising in the 2EWKM:

\begin{equation}
\centering
Q(t) = \frac{P(t)}{R_{out}} +C_{out}\frac{dP(t)}{dt},
\label{eq:eq6}
\end{equation}

\noindent where $Q$ is the descending aorta flow rate in $\si{m^3/s}$, $P$ is the pressure in $\si{mmHg}$, $R_{out}$ is the distal resistance and $C_{out}$ is the capacitance accounting for the vessel compliance.

Given a value for diastolic blood pressure (typically $80\si{mmHg}$), the equation above becomes a well-posed initial value problem, where values of $R_{out}$ and $C_{out}$ could be found such that the time integration of the Eq. \eqref{eq:eq6} results in a periodic pressure waveform. For the idealized case, a flow waveform with a sinosuidal systolic part and a null diastolic part is used,  which allows an analytical solution for pressure can be obtained \cite{catanho2012model}.

\subsection{High-performance computing with GPUs}
\label{secparallel}

Detailed numerical simulations of turbulent flows are costly in nature due to the required large grid resolutions \cite{pope2000turbulent}.  For instance,  using a resolution of $512\times512\times512$ grid points integrated in time for an average of $2.7 \times 10^4$ time-steps,  our massively parallel CPU-based solver \cite{henniger2010high} would require approximately two months to complete a single-beat aortic simulation using only CPU cores of 8 hybrid nodes (8$\times$12 Intel Xeon E5-2690 v3 cores,  $8\times64$GB RAM) of a Tier 0 supercomputer ($Piz~Daint$,  Swiss National Supercomputing Centre).  Resorting to our GPU-based implementation brings down this simulation time to 3 days,  which is more practical in a clinical environment.  Resorting to a higher number of GPUs,  even though may be accessible only through cloud based clusters,  would lead to even better solver throughput.  For instance resorting to 256 P100 GPUs would bring down the simulation time to 1-2hrs.  The reduction of computational time at the same rate as the increasing of computational resources is not the case for majority of parallel implementations due communication overhead and data transfer latency,  nevertheless, it is achieved here thanks to strong scaling capability of current GPU implementation \cite{zolfaghari2021high}.

Parallelism is achieved in a two-step process.  First the computational domain is decomposed into blocks of equal sizes, which are then processed in parallel using the message passing interface (MPI) \cite{henniger2010high,zolfaghari2019high}, and then each MPI block is decomposed into its grid points, and each grid point is then processed using a CUDA (compute unified device architecture) thread \cite{zolfaghari2021high,zolfaghari2019high}.  The solver uses a novel high-throughput Poisson solver for pressure which significantly reduces the time-to-solution,  resulting in speedups of factor 10-300 compared to a parallel CPU-based solver \cite{zolfaghari2021high}.  Even though the original solver was designed based on periodic boundary conditions, it has been modified here for incorporation of Dirichlet type pressure outflow boundary conditions.




All cases presented in the paper have been run using a simulation box resolution of $512\times512\times512$.  The simulation runs for a physical time of one heart beat have been performed for roughly three days using 8 NVIDIA P100 GPUs, each seeded by 8 CPU cores. The average time-step for the simulations was $dt = 0.000037\si{s}$.  A CFL number of 0.5 was set. All runs were performed on Haswell nodes of Cray XC40/50 $Piz~Daint$ supercomputer.

\section{Results and Discussion}
\label{sec:MHVDNS}
\subsection{Hemodynamics of the AS and virtually restored scenarios }

\subsubsection{Overview of the patient specific and idealized cases}
\label{geometries}

\begin{figure}
\centering
\includegraphics[width=1\textwidth]{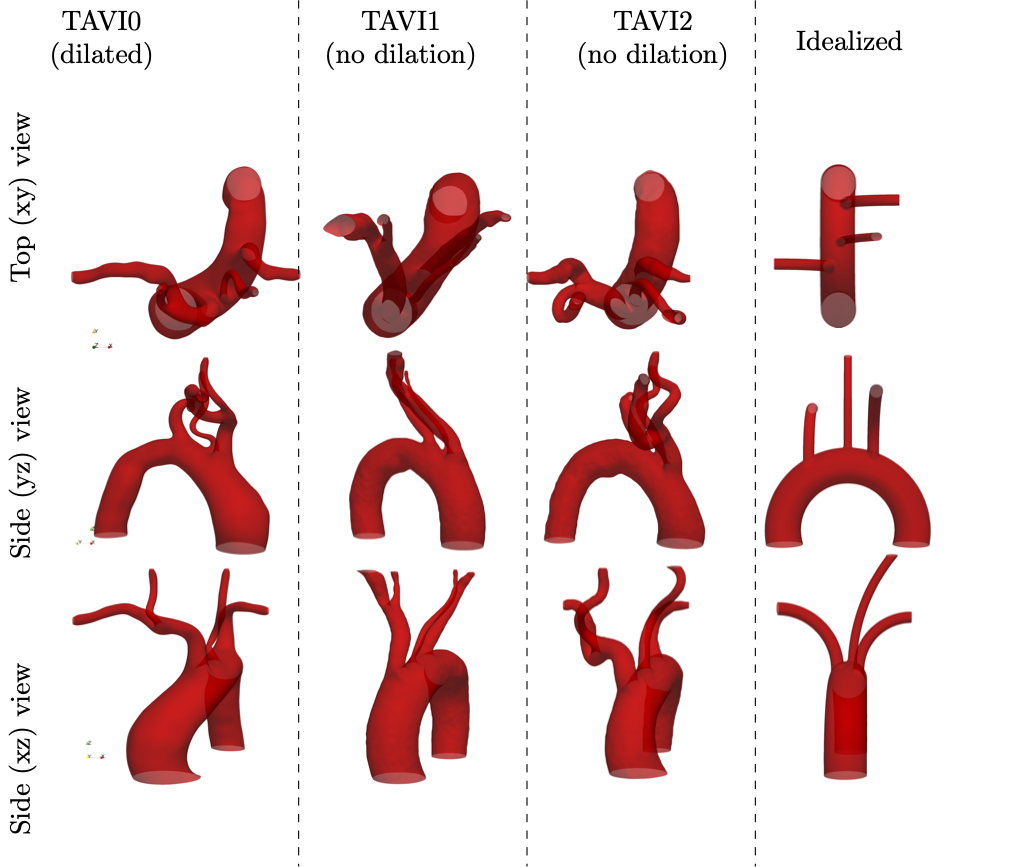}
\caption{Geometric models for patients \enquote{TAVI0}, \enquote{TAVI1} and \enquote{TAVI2},  as well as the Idealized model \enquote{Idealized}.}
\label{fig:all_geometries}
\end{figure}

The reconstructed geometrical models for AS cases,  labelled as TAVI0, TAVI1, TAVI2 (see Table \ref{tab:dimsCT}) and an idealized case are shown in figure \ref{fig:all_geometries}.  The TAVI0 case has marked dilation (AAo diameter of 4cm),  while cases TAVI1 and TAVI2 do not present dilation.  The figure shows the geometries sorted in terms of out-of-plane bending of the aortic arch, where TAVI0 has the highest bending,  followed by TAVI1, TAVI2 and Idealized (which has no out-of-plane bending) cases.  In addition to diseased cases TAVI0-2,  we also investigate the hemodynamics in a restored (i.e.  after the narrowed valve is replaced with a valve prosthesis) condition for the TAVI0 case.  We model the restored flow by increasing the aortic jet area three times as that in the diseased case, while maintaining the flow rate waveform.  This simplified model corresponds to a reduction in peak flow velocity from 2.7$\si{m/s}$ to $0.9\si{m/s}$, which is in the healthy range.  Further to a baseline restored case which we label as \enquote{TAVI0-Restored},  we also investigate a case of restored flow under extreme stress (labelled as \enquote{TAVI0-Restored+S}),  which has the same aortic valve jet area as the \enquote{Restored} case,  but a peak velocity twice that case (i.e.  $1.8\si{m/s}$).

\begin{table}[htb]
    \caption{Physical dimensions of the geometric and waveform input models.  The \enquote{TAVI0 Restored} and \enquote{TAVI0-Restored+S} (which is the stressed version of restored case) cases are based on the geometry of TAVI0 case.  The inflow and out flow cross sections are given under \enquote{D- Slice} (datum slice,  which is the slice at lowest $z$ inside $\Omega_{CT}$ subdomain which is introduced in figure \ref{fig:fig2}).  Lower curve shows the inlet and the upper curve shows the outlet boundary) and are roughly of a egg-like oval shape.  The sizes of these inflow and outflow boundaries given in a $a\times b$ format, where $a$ is the major axis and $b$ is the minor axis of the oval.}
    \label{tab:dimsCT}
    \centering
        \begin{tabular}{c | S[table-format=2.2]|
                       *{3}{S[table-format=2.2]|}
                             S[table-format=2.2]}
        \hline
        Case
            & {\makecell{CT Resolution \\ $(\si{mm} \times \si{mm} \times \si{mm})$}}                              
                        & {\makecell{ Inflow Size \\ $(\si{mm}\times\si{mm})$}}        
                         & {\makecell{ Outflow Size \\ $(\si{mm}\times\si{mm})$}}   
                              & {\makecell{ AV jet radius \\ $(\si{mm})$}}                         
                         & {\makecell{D- Slice}}                               \\  \hline
        TAVI0 &   {$0.4 \times 0.4 \times 0.5$}  & {$40\times 36$} & {$28\times 28$} & 7.0 & \raisebox{-0.5\totalheight}{\includegraphics[scale=0.03]{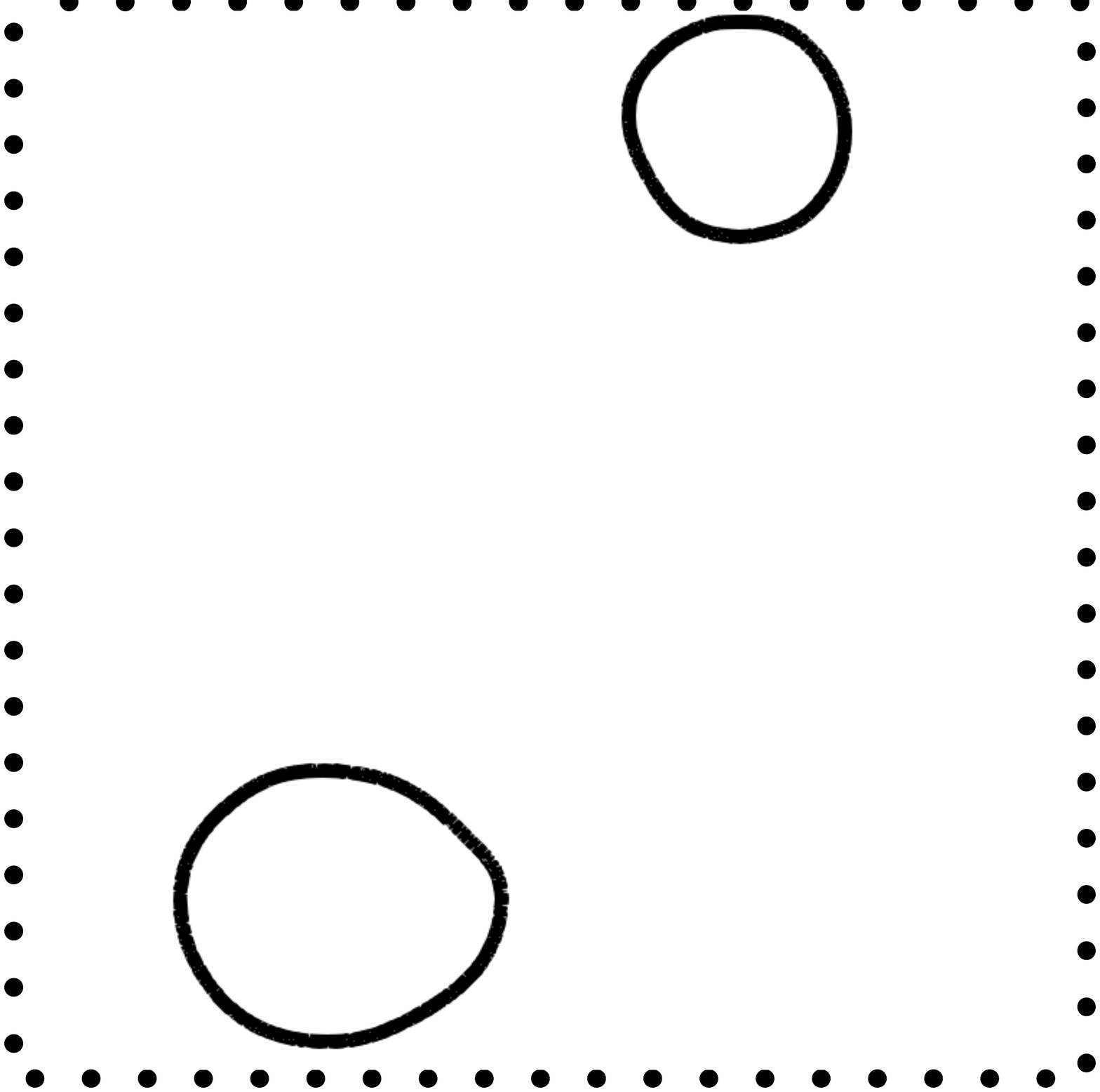}}   \\ \hline
        TAVI1 & { $0.746\times0.746\times0.75$}   & {$30\times 28$} & {$26\times 24$} & 7.7 & \raisebox{-0.5\totalheight}{\includegraphics[scale=0.03]{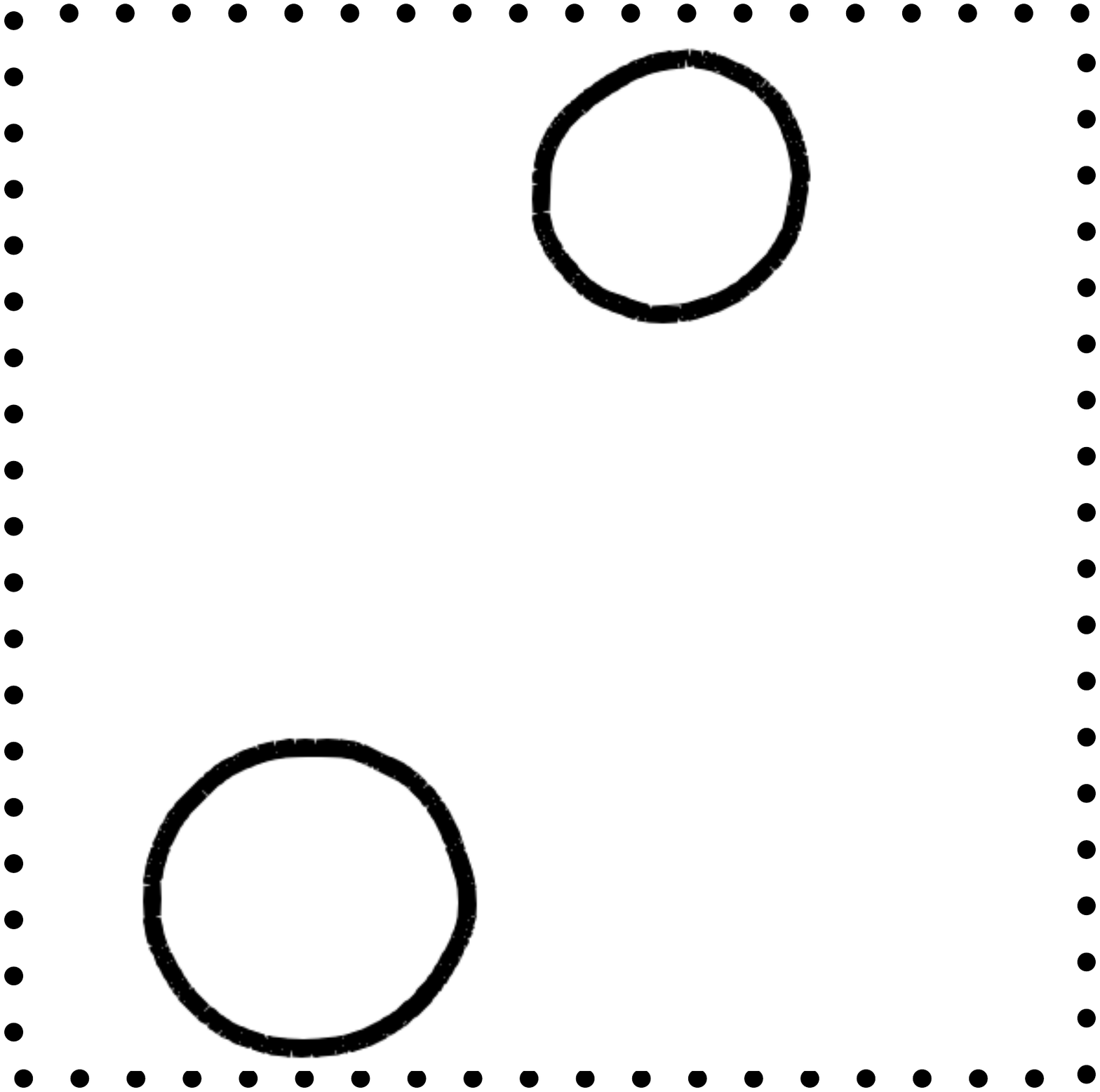}}   \\ \hline
        TAVI2 & { $0.779\times0.779\times0.75$}  & {$35 \times 31$} & {$26\times 26$} & 7.5 & \raisebox{-0.5\totalheight}{\includegraphics[scale=0.03]{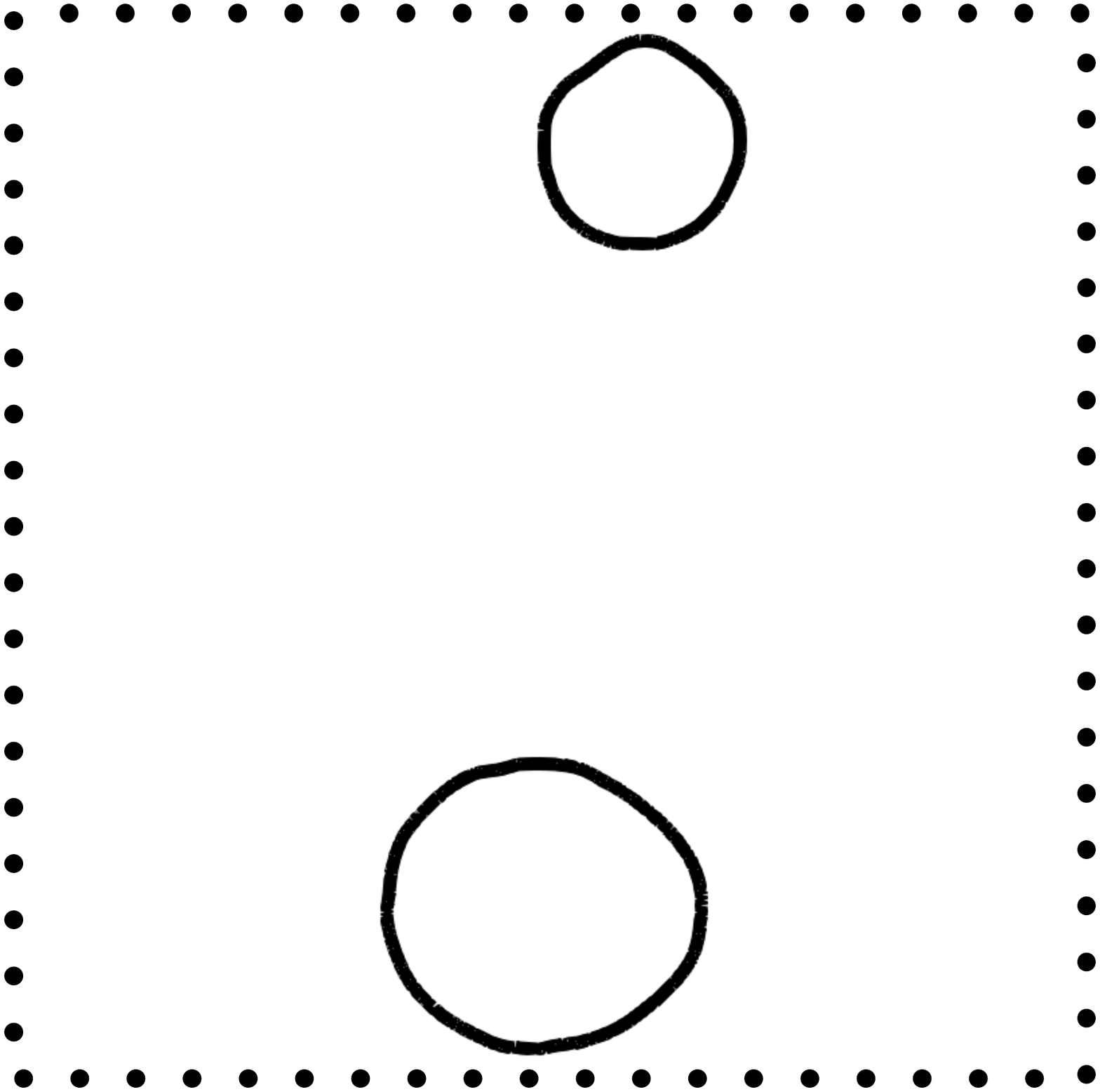}} \\ \hline
        Idealized & NA  & {$29 \times 29$} & {$29 \times 29$} & 6.8 & \raisebox{-0.5\totalheight}{\includegraphics[scale=0.03]{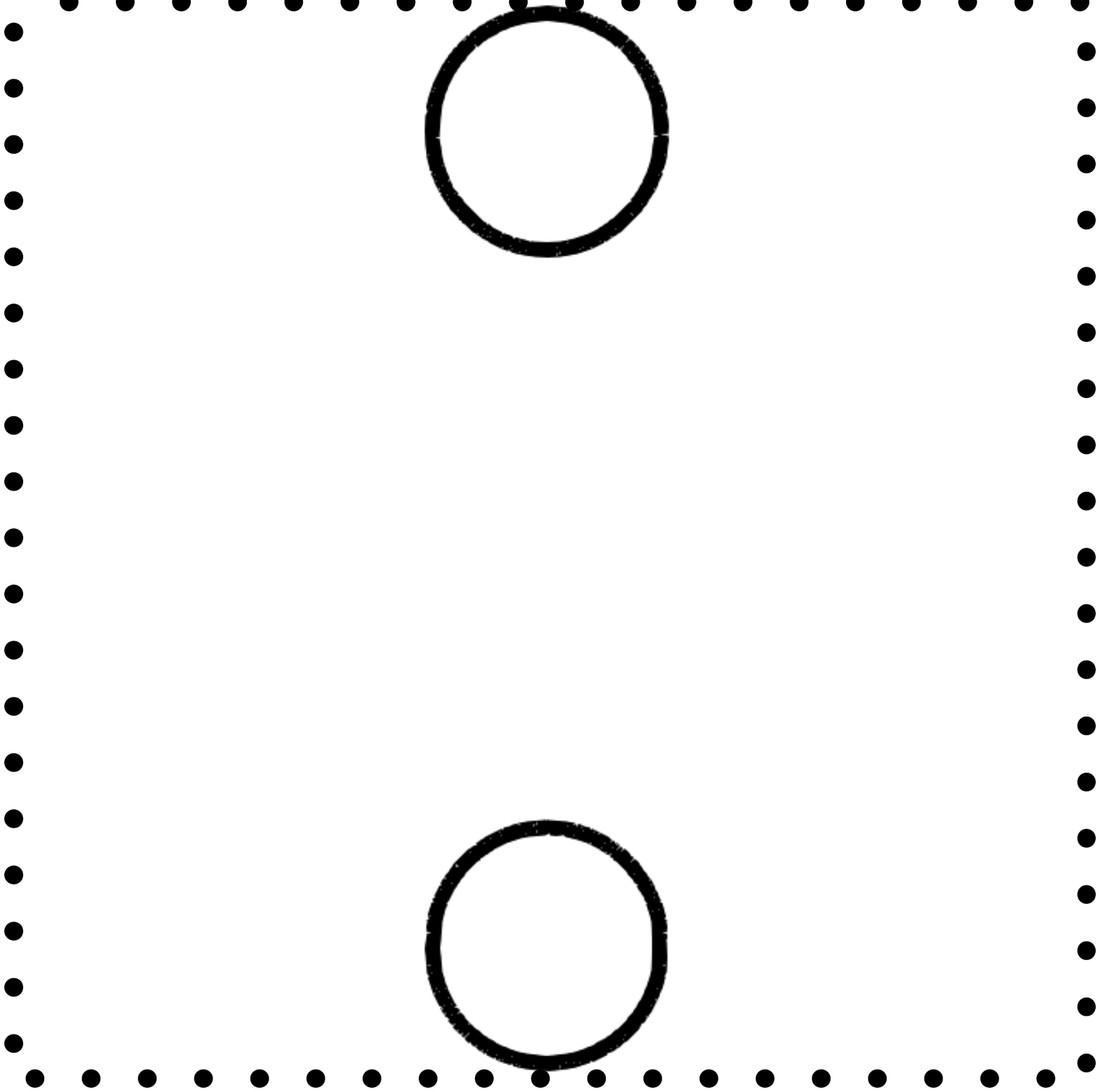}}  \\ \hline
        TAVI0 Restored &   {$0.4 \times 0.4 \times 0.5$}  & {$40\times 36$}  & {$28\times 28$} & 10.2 &  \raisebox{-0.5\totalheight}{\includegraphics[scale=0.03]{media/intensity_slice/TAVI000_2}} \\ \hline
        TAVI0-Restored+S &   {$0.4 \times 0.4 \times 0.5$}  & {$40\times 36$}  & {$28\times 28$} & 10.2 & \raisebox{-0.5\totalheight}{\includegraphics[scale=0.03]{media/intensity_slice/TAVI000_2}} \\ \hline

        \end{tabular}
\end{table}

\subsubsection{Wall shear stress: from waves to a chaotic character}

\begin{figure}
\centering
\includegraphics[width=1\textwidth]{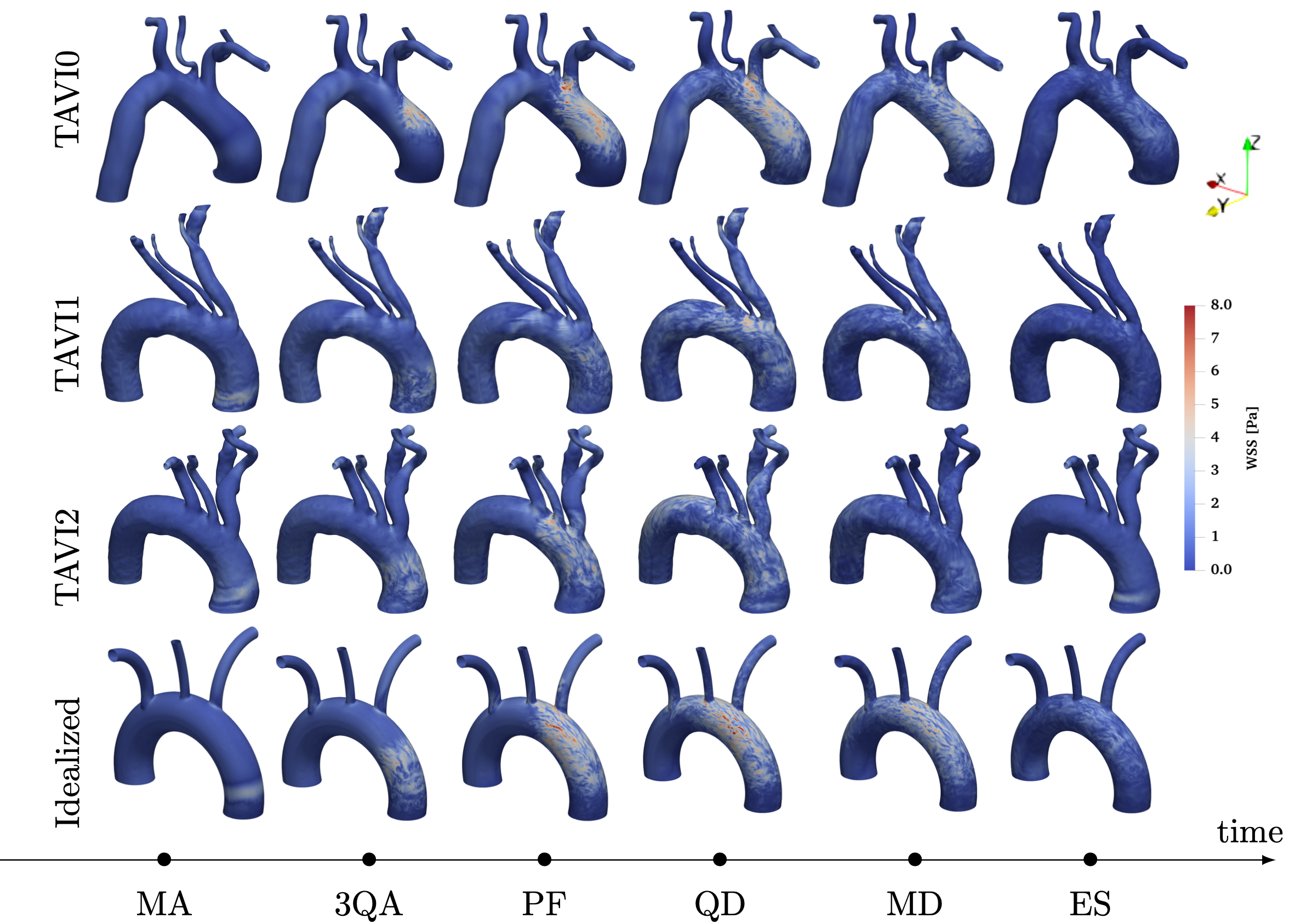}
\caption{Wall shear stress magnitude maps for instances MA,  3QA, PF, QD,  MD and ES are shown for patient-specific cases TAVI0, TAVI1 and TAVI2 as well as the Idealized case.  A persistent focal area of elevated and oscillatory wall shear stress is observed on the outer curvature of the ascending aorta for TAVI0 and the Idealized cases,  whereas elevated values of WSS are more dispersed in space for cases TAVI1 and TAVI2.  This could be due to more downstream branching location of the Brachiocephalic artery for the TAVI0 and Idealized cases, compared to TAVI1 and TAVI2.  The former cases allow jet impingement on the vessel wall without further disturbance due to cross-flow caused by this neck artery,  while the latter cases include further disturbing of this zone by this cross-flow effect (which acts as a source of blowing/suction of fluid).  }
\label{fig:wss_all}
\end{figure}

\begin{figure}
\centering
\includegraphics[width=0.5\textwidth]{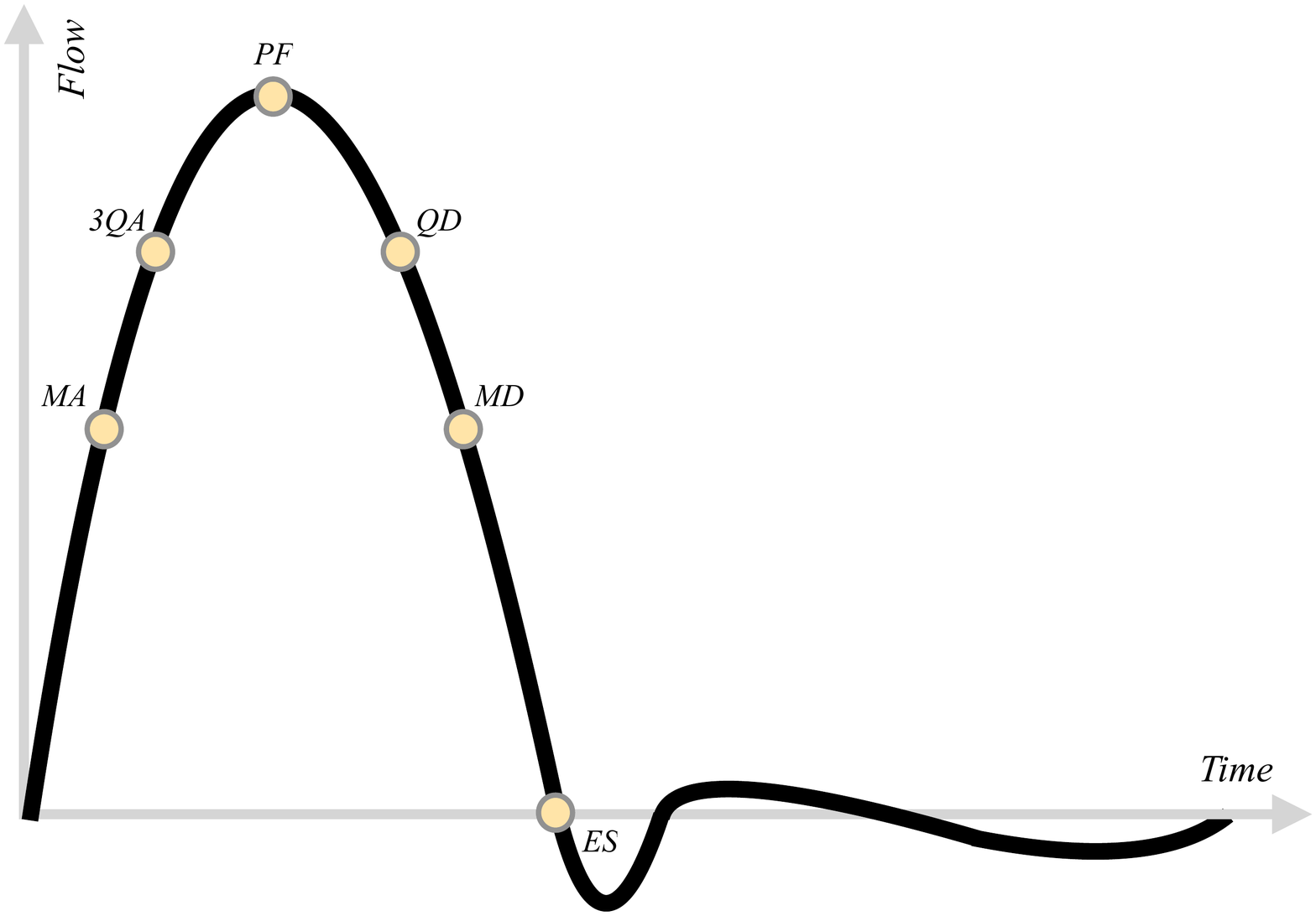}
\caption{Time instance references used in the paper for the in-beat snapshot presentations:  peak flow (PF) denotes when flow rate reaches its maximum.  End systole (ES), denotes the state of flow at the end of systolic deceleration.  Other instances are then defined accordingly as follows: mid-acceleration (MA) is temporally halfway between the start of the pulse ($t=0$) and PF,  three-quarters acceleration (3QA) is halfway between MA and PF,  mid-deceleration (MD) is halfway between PF and ES,  quarter-deceleration (QD) is halfway between PF and MD. }
\label{fig:instances_legend}
\end{figure}

The magnitude of instantaneous wall shear stress (WSS) is calculated as 

\begin{equation}
{\sigma}_w = | {(\pmb{\sigma}\cdot \mathbf{n}) - (\pmb{\sigma}\cdot\mathbf{n}\cdot\mathbf{n})\mathbf{n}}|
\end{equation}

\noindent where $\pmb{\sigma}$ is the fluid's Cauchy stress tensor and $\mathbf{n}$ is the outward normal to the aortic wall.  Figure \ref{fig:wss_all} shows instantaneous maps of WSS for cases TAVI0, TAVI1, TAVI2 and Idealized which are demonstrated for six time instances, namely,  MA (mid-acceleration), 3QA (three-quarters acceleration), PF (peak flow), QD (quarter deceleration), MD (mid-deceleration) and ES (end systole).  These time instances within a standard velocity waveform are indicated in figure \ref{fig:instances_legend}. It is seen that a focal area of spatially oscillatory and elevated WSS exists on the outer curvature of the AAo for the TAVI0 cases.  Cases TAVI1 and TAVI2, on the other hand, show elevated WSS  values more dispersed in space,  i.e.,  more extended both in the circumferential direction and also towards the distal aortic arch.  The focal zone of high WSS,  which emerges around PF and persists through the beat cycle roughly until MD,  is suspected to form due to the turbulent impingement of the strong stenotic jet on the outer wall of the ascending aorta (a similar case for a tricuspid AS patient with dilation has been shown in \cite{manchester2021analysis}).  This focal area has been shown for patients with bicuspic aortic valve (BAV),  and extensively addressed in the context of regional aortic remodelling in the recent literature \cite{pasta2017silico, guzzardi2015valve, soulat2022association} (see especially  \cite{guzzardi2015valve}, where the regional zone of elevated WSS was illustrated to be linked to regional aortic remodelling).

\begin{figure}
\centering
\includegraphics[width=1\textwidth]{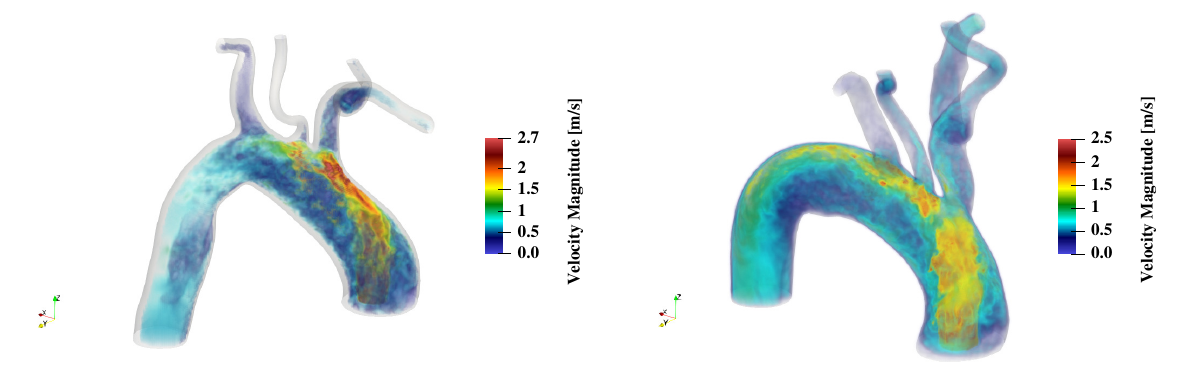}
\caption{Volume rendering of the undisturbed (left, TAVI0 case) and disturbed (right,  TAVI2) aortic jet impingement on the aortic wall at time QD.  The undisturbed impingement case shows that the aortic jet changes direction without breakdown while the disturbed impingement case shows the jet undergoing large disturbances while impinging (the destabilizing effect of blowing/suction by Brachiocephalic artery).}
\label{fig:volume_rendering}
\end{figure}

 The absence of such focal zone for the cases TAVI1-2 despite aortic jet velocities on par with the TAVI0 case is likely to be due to the secondary disturbing effect of the Brachiocephalic artery inlet through coincidence with the impingement zone.   As the middle row (yz view) of figure \ref{fig:all_geometries} suggests, the Brachiocephalic inlet is slightly more downstream for TAVI0 than TAVI1 and TAVI2; this leaves more available space for the incoming jet to impinge without further disturbance.  However,  presence of the brachiocephalic artery inlet at the jet impingement site mimics wall suction mechanism which could disrupt its localized impact (see figure \ref{fig:volume_rendering}),  while it could enhance or weaken the background turbulent flow. 

The hypothesis that the focal zone of high WSS occurs when brachiocephalic artery inlet is sufficiently distal to the impingement zone is tested using the Idealized model presented in figure \ref{fig:all_geometries}.  The wall shear stress snapshots for this case are shown next to patient cases in figure \ref{fig:wss_all}. It is shown that the focal zone is present throughout the systolic phase which is in agreement with our hypothesis.

%

\begin{figure}
\centering
\includegraphics[width=1\textwidth]{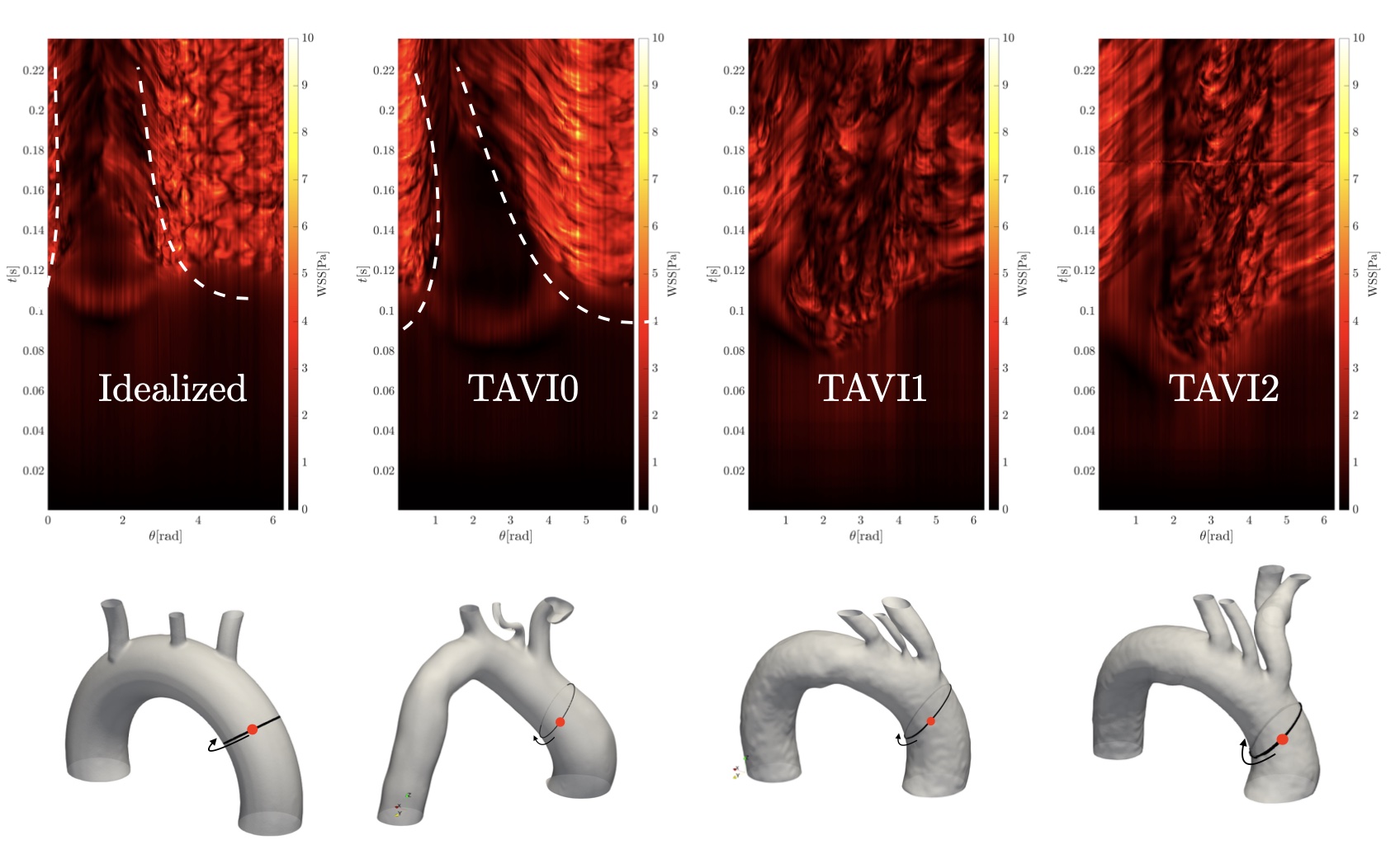}
\caption {Evolution of wall shear stress values over time (top panels) for a band (solid black line shown on the bottom panels) in the ascending aortic region.  For each point on the band, the angle $\theta$ is defined with respect to an origin on the surface enclosed within the band,  and is taken to be the mean value of $x$, $y$ and $z$ coordinates of the surface.  Red bullet mark shows $\theta =0$ and the positive azimuthal direction is shown by arrows.  Wall shear stress data on the band are then collected in time and plotted over a period from the start of the pulse $t=0$ until $t =0.225\si{s}$, which is slightly past the peak flow.  It can be seen that wall shear stress peaks cover the entire band for cases TAVI1 and TAVI2, but are concentrated on one side of the band for cases TAVI0 and Idealized. The white dashed lines show approximately the boundary of high wall shear stress zones and the quiet zone on for the cases TAVI0 and Idealized. }
\label{fig:theta_time}
\end{figure}

Figure \ref{fig:theta_time} shows the separation in WSS patterns for cases with focal elevation (TAVI0 and Idealized) next to that with a dispersed elevation (TAVI1-2). Time histories of WSS data taken at a closed band around the ascending aorta artery inlet are plotted over the systolic acceleration phase.   A focal but more intense zone of WSS for TAVI0 and Idealized cases are observed, whereas, WSS oscillations span the entire azimuthal extent of the closed band for cases TAVI1 and TAVI2.

The mere locality of the elevated WSS in the ascending aorta,  and even its slightly higher magnitude, do not justify the dilation of the TAVI0 case as opposed to cases TAVI1-2,  because the wall shear stress is elevated for all cases in AAo anyway.  The flow deceleration around the jet impingement zone however is expected to develop a zone of high pressure,  which could be a driving force for dilation.  We emphasize again that the above observations are made under several assumptions and could be overruled by anatomical factors of the wall tissue.  Our goal is to present the hemodynamic differences over this given spectrum of aortic geometries, which could help understand a possible contribution of hemodynamic factors in the aortic dilation.In the next section, we take advantage of geometrical symmetry of the Idealized case and investigate the possibility of a high pressure pocket underneath the impingement zone.

\subsubsection{Pressure behaviour in disturbed vs undisturbed  stenotic jet impingement}
\label{sec:pre_and_vel}
\begin{figure}
\centering
\includegraphics[width=1\textwidth]{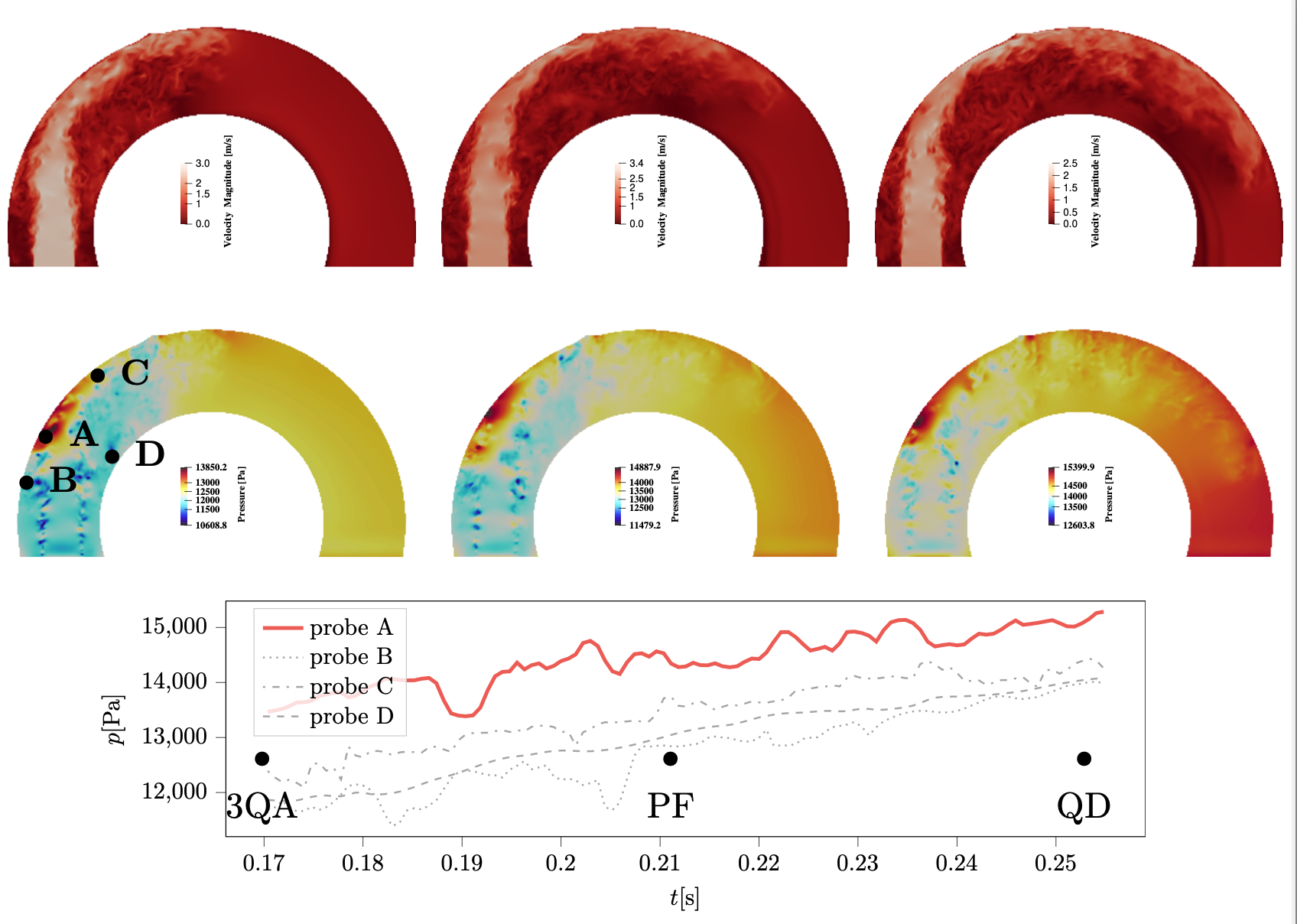}
%
%
%
%
%
%
%
\caption{Realization of the jet impingement process through velocity magnitude (top panels) and pressure (middle panels) visualizations in the Idealized case.  From left to right,  time instances 3QA, PF and QD are shown.  The bottom plot shows the time history of pressure values taken on four probes A, B, C and D (see the middle left panel) around the impingement zone.  It is shown that probe A which is located roughly close to the centre of impingement zone, consistently marks a higher pressure (of 1000Pa approximately) than other neighbouring probes.  }
\label{fig:impingement_idealized}
\end{figure}

Figure \ref{fig:impingement_idealized} shows the jet impingement on the aortic surface for the Idealized case. Three snapshots of velocity magnitude are taken on a slice cutting through the aortic geometry on its axis of symmetry normal to the $x$ axis. They show the jet velocity changes due to impingement at outer AAo wall due to deceleration.  This flow deceleration corresponds to a high pressure pocket forming at the impingement zone (cf. middle rows of figure \ref{fig:impingement_idealized}). Continuous pressure signals between this three snapshots and at four selected probes A, B, C, and D (shown on the middle left panel of the figure) confirm that the pressure at the centre of impingement (point A) remains consistently higher than the other neighbouring probes B, C and D. A mean pressure difference of approximately $1000\si{Pa}$ focused on a single location can have implications on local dilation of the vessel, which may have been deteriorated by the high shear stress.  This focal zone of high pressure acts as the normal force required for the local dilation of the aorta.  In this sense,  even though elevated and oscillatory wall shear stresses weaken the arterial wall for all cases (maybe more locally for the dilated case),  the elevated pressure, if persistent in time,  could drive the expansion process by forcing the weakened wall matrix in the outward normal direction (i.e., it acts as tensile stress acting to expand the aorta).

\begin{figure}
\centering
\includegraphics[width=1\textwidth]{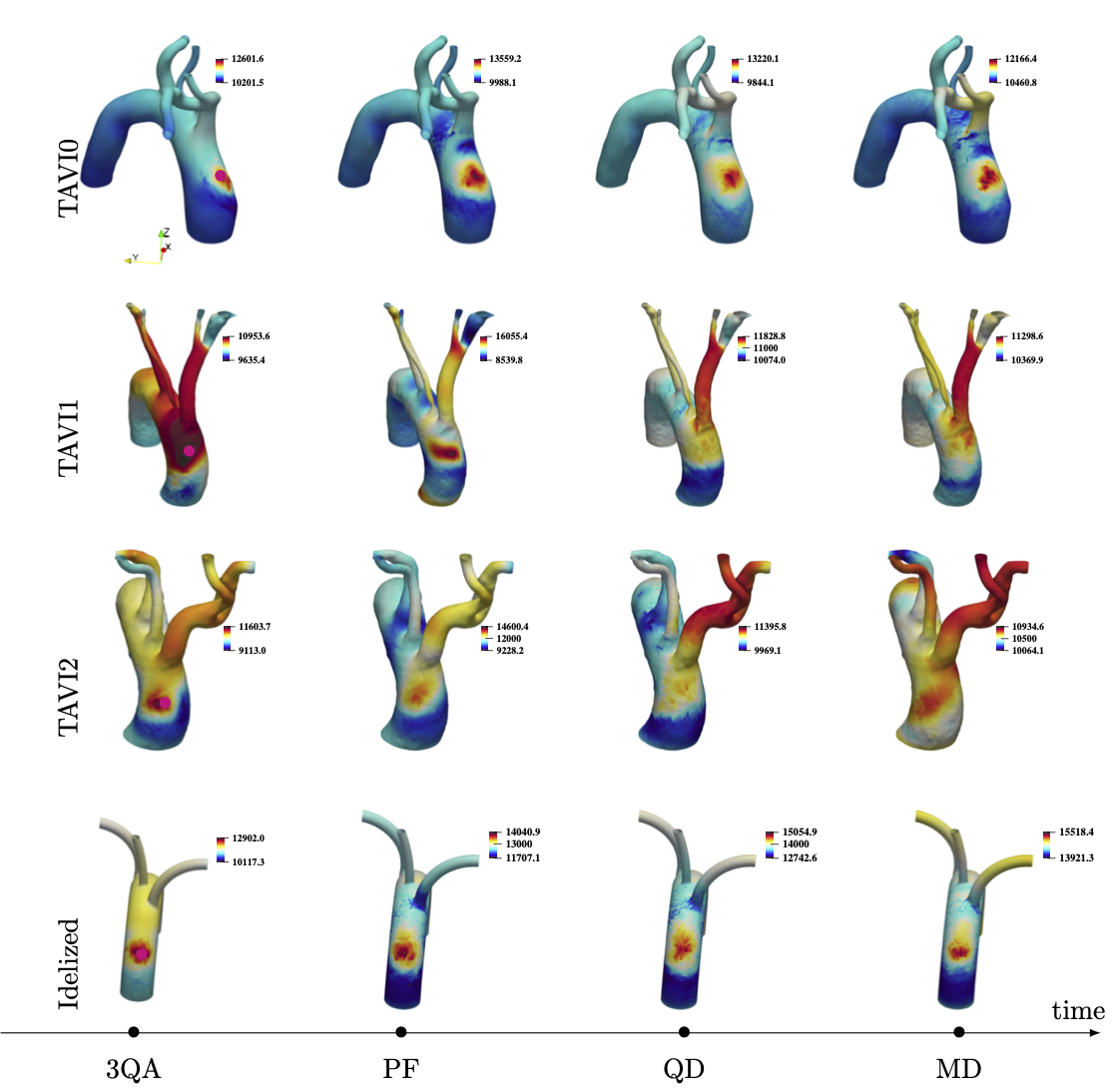}
\caption{Wall pressure maps for instances 3QA, PF, QD,  MD are shown for patient-specific cases TAVI0, TAVI1 and TAVI2 as well as the Idealized case Idealized.  A persistent zone of high pressure, which corresponds to a high wall shear stress zone (cf. figure \ref{fig:wss_all}), is observed on the ascending aorta for cases TAVI0 and Idealized,  but not for the cases TAVI1 and TAVI2.  }
\label{fig:all_pressure}
\end{figure}

\begin{figure}
\centering
\includegraphics[width=0.7\textwidth]{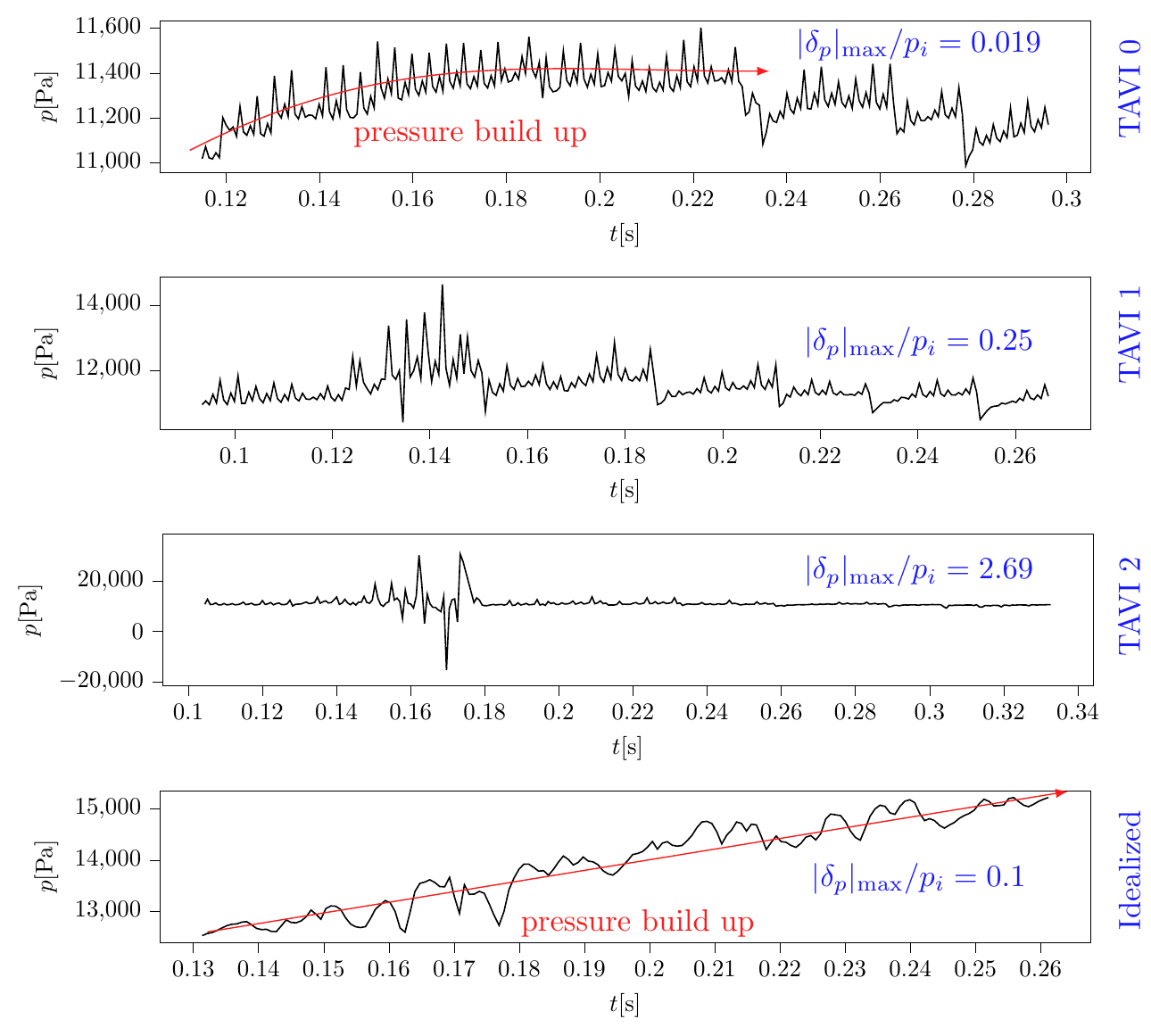}
%
\caption{Pressure signal taken as a time series of pressure at a probe located on the aortic wall slightly upstream of neck arteries (the probe is marked in magenta on the left column of figure \ref{fig:all_pressure},  and is located at centre of the HWSS zone) for TAVI0, TAVI 1, TAVI 2, and Idealized cases. Cases TAVI 0 and Idealized show distinct pressure build up. Cases TAVI 1 and TAVI 2 lack a pressure build up, but show strong oscillations near peak flow which is shown by the dimenstionless parameter $|\delta_{p}|_{\max}/p_{i}$, where $\delta_{p}$ denotes twice the maximum amplitude of the oscillations, and $p_i$ is the initial pressure value taken at 3QA instance.}
\label{fig:pressure_signals}
\end{figure}

Figure \ref{fig:all_pressure} shows systolic snapshots of wall pressure for TAVI0-2 and Idealized cases. Consistent to the WSS interpretations made in the previous section, cases TAVI0 and Idealized present regional focal zones of high wall pressure, which persist until mid-deceleration (MD) instance of the beat cycle.  Cases TAVI1 and TAVI2 show no clear regional pressure peak but rather an oscillatory pressure behaviour.  A frequently sampled signal of wall pressure (sampling rate of 1400Hz), measured on the exterior surface of the ascneding aorta further confirms this observation (figure \ref{fig:pressure_signals}). Pressure signals for TAVI0 and Idealized cases show a \enquote{pressure buildup} effect (marked in red) with only low amplitude oscillations, while the pressure signals for TAVI1 and TAVI2 cases show rather higher amplitude oscillations near peak flow without any apparent pressure buildup.  This stronger oscillations show that the suction effect of the brachiocephalic artery has led to an even higher amplitude oscillations for the TAVI1 and TAVI2 cases (in terms of WSS, this was seen as a more extended wall footprint, yet with a lower magnitude than the impingement case).  A measure for the relative amplitude of oscillations with respect to the background is given in figure \ref{fig:pressure_signals} as $|\delta_{p}|_{\max}/p_{i}$, where $|\delta p|$ is twice the amplitude of oscillations present in the buildup part of the signal,  and $p_i$ is the pressure at the onset of pressure buildup.  Large pressure values observed in cases TAVI1-2 could be alarming in terms of enhancing the risk of a sudden aortic dissection,  however,  the strongly oscillatory behaviour as opposed to a persistent buildup may explain why these cases do not undergo dilation despite turbulence.

\subsection{Hemodynamics of the restored flow in the dilated aorta}

Treatment scenarios for AS patients with aortic dilation may include the aortic root replacement next to the aortic valve replacement (AVR).  This is justified by the increased risk of future aortic dissection  in presence of dilation,  which is a marker of arterial wall deterioration.  Even though the progression and rapture of a dilated aorta is a biologically involved process,  hemodynamics may also play a role in this phenomenon. To this end,  we study the changes in wall shear stress and pressure in case of a restored aortic flow in the dilated case TAVI0.  The restored cases labelled as TAVI0-Restored (without stress) and TAVI0-Restored+S (with extreme stress) are simulated.   Detailed specifications of these cases can be found in section \ref{geometries}.    

\begin{figure}
\centering
\includegraphics[width=0.7\textwidth]{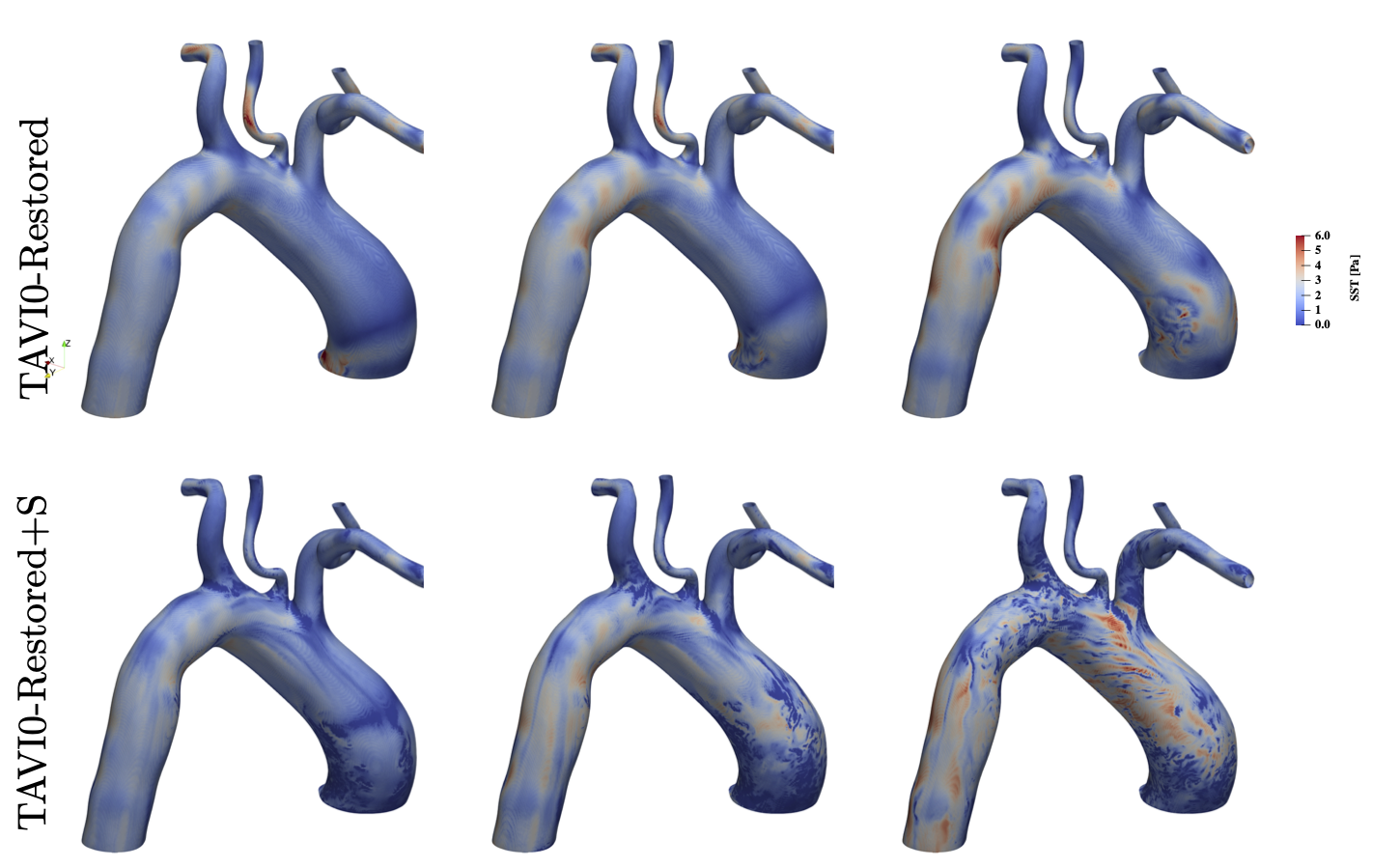}
%
\caption{Wall shear stress maps for the virtually restored case of TAVI0 at times 3QA,  PF and QD.  Top row corresponds to restored flow with the same cardiac output ({no stress} scenario) as the stenosis case, and the bottom row corresponds to a {stressed} scenario where cardiac output is doubled.  It is seen that both scenarios do not mark a high wall shear stress zone on the ascending aorta as opposed to the stenosis case.  However,  new zones of high wall shear stress have emerged downstream of the aortic arch bend towards the descending aorta.  Signs of helical wall-shear stress waves can be also observed on the ascending aorta for the {stressed} scenario (see bottom right panel).}
\label{fig:wss_restored}
\end{figure}

\begin{figure}
\centering
\includegraphics[width=0.7\textwidth]{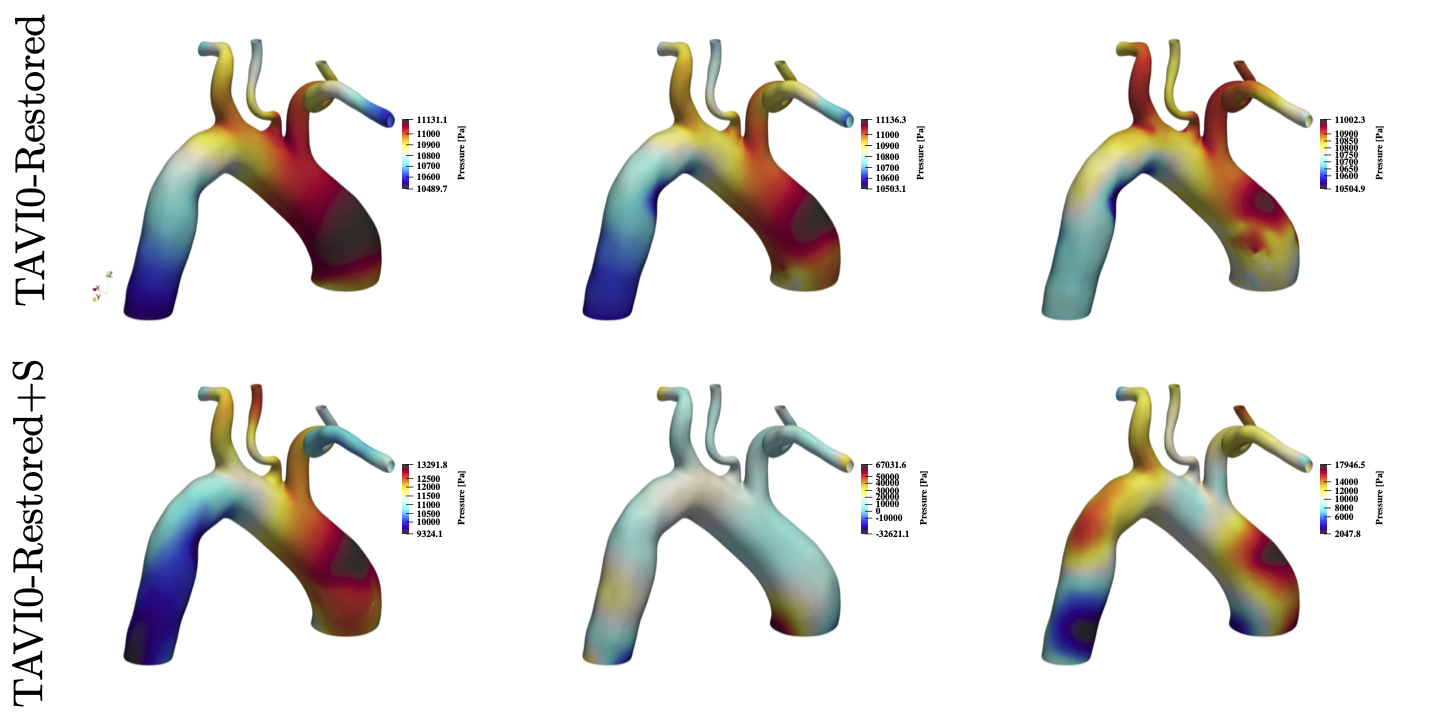}
%
%
\caption{Wall pressure maps for the virtually restored case of TAVI0 at times 3QA,  PF and QD.  Top row corresponds to restored flow with the same cardiac output ({unstressed} scenario) as the stenosis case, and the bottom row corresponds to a {stressed} scenario where cardiac output is doubled.  It is seen that the regional zone of high wall pressure on the ascending aorta which was observed in the stenosis case is no longer present.  The unstressed scenario shows an almost uniformly decaying pressure from ascending towards the descending aorta, but the stressed scenario shows an oscillatory behaviour,  which is likely to be related to the helical wave footprints on wall shear stress maps for this case (see figure \ref{fig:wss_restored}). }
\label{fig:pressure_restored}
\end{figure}

Figure \ref{fig:wss_restored} shows the wall shear stress distributions for the TAVI0-Restored and TAVI0-Restored+S flow scenarios at time instances 3QA, PF and QD.   It is seen that the focal zone of elevated WSS has been removed from the restored and unstressed cases.  The wall shear stress levels are in general half that of the corresponding diseased (TAVI0) case.  The stressed case (bottom panels of the figure) does not indicate any sign of the focal zone seen for the TAVI0 case,  nevertheless,  this case exhibits elevated levels of WSS throughout the aortic wall.  Figure \ref{fig:pressure_restored} shows the pressure maps of cases TAVI0-Restored and TAVI0-Restored+S for times 3QA,  PF and QD.  Consistent with the wall shear stress outcome (figure \ref{fig:wss_restored}), the regional pressure peak has been removed for the restored case at both stressed and unstressed conditions.  The unstressed case TAVI0-Restored shows a uniformly decaying pressure from ascending to the descending aorta,  while the stressed case shows an oscillatory behaviour, although it does not demonstrate any persistent regional pressure peak.

These observations suggest that the dilated aorta does not in itself lead to a disorganized flow in the ascending aorta (the restored flow without stress does not exhibit any chaos in AAo).   Finally,  figure \ref{fig:wss_restored} also shows that the maximum wall shear stresses of up to 6Pa are found in the distal arch area for TAVI0-Restored case.  This zone of high WSS did not appear prior to AVR (see TAVI0 case in figure \ref{fig:wss_all}),  and may be related to dilation or other anatomical features of TAVI0 case.  The latter mechanism is not addressed in this paper, however,  the velocity field underneath this emerging zone of elevated WSS post AVR are discussed in the following.

\begin{figure}
\centering
\includegraphics[width=0.5\textwidth]{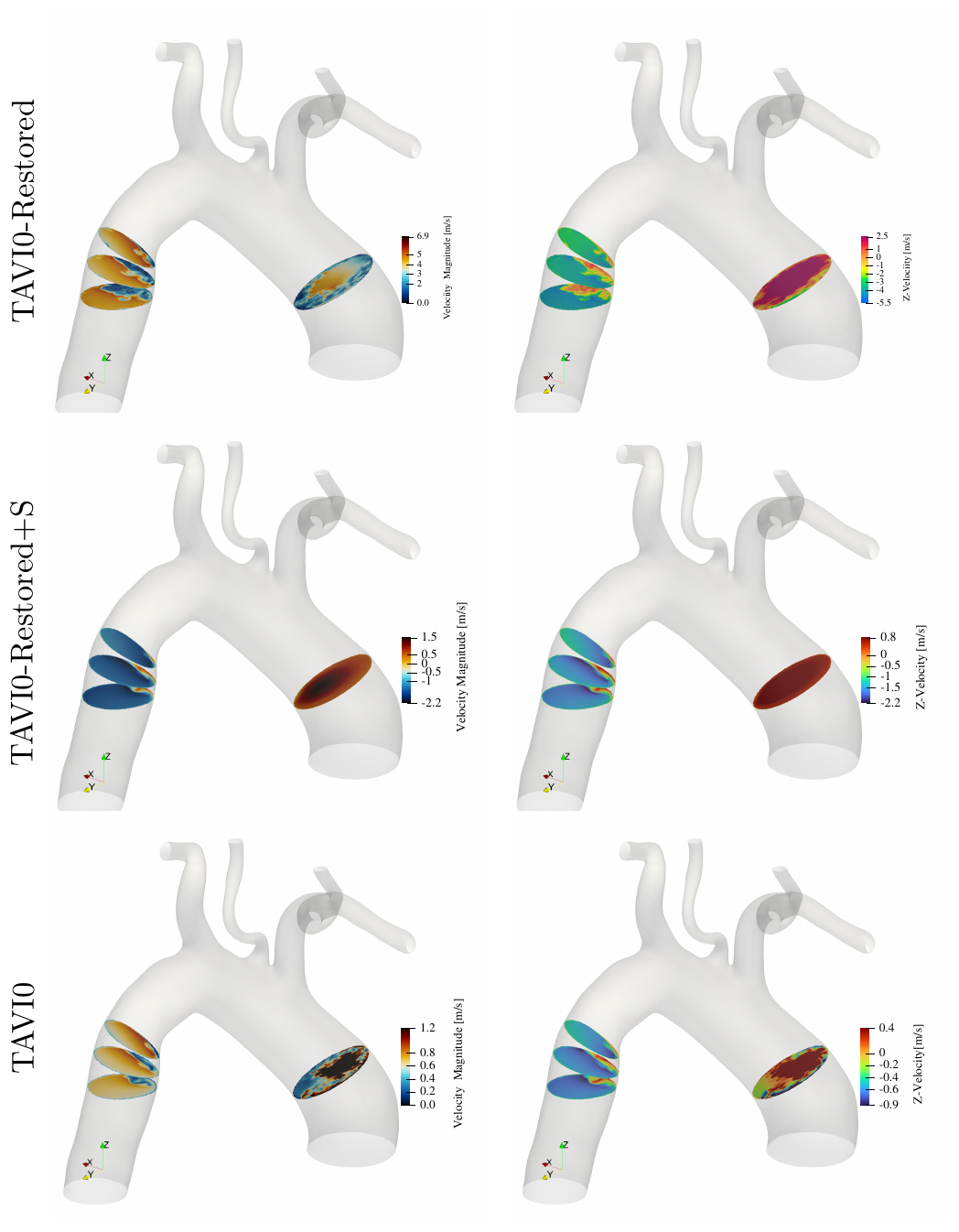}
%
\caption{Peak flow velocity magnitude (left) and $z-$component of the velocity field (right) for the TAVI0 case at stenosis (bottom), restored and unstressed (middle) and restored and stressed (top) cases. Even though the flow is less chaotic after restoration in the ascending aorta, the descending aortic flow marks a major instability (a separation bubble) which is linked to elevated wall shear stress in that location (see figure \ref{fig:wss_restored}). This bubble could be due to the major out-of-plane bending of the aorta. }
\label{fig:velocities_restored_vs_stenosis}
\end{figure}

Figure \ref{fig:velocities_restored_vs_stenosis} shows four slices of the background velocity fields at peak flow for the restored scenarios next to the stenosis case.  It is seen that the TAVI0 case undergoes severe turbulence in the ascending aorta,  whereas the flow in the descending aorta remains largely laminar (only a tiny region of reverse flow is observed).  The laminar flow in this zone is consistent with the lower wall shear stress observations in figure \ref{fig:wss_all}.  For the TAVI0-Restored case (middle row panels of figure \ref{fig:velocities_restored_vs_stenosis} ),  the flow throughout the AAo becomes laminar,  which also corresponds to lower wall shear stress levels than the diseased case (see figures \ref{fig:wss_all} and  \ref{fig:wss_restored}).  However,  this restored case indicates a zone of elevated WSS in the distal arch area as discussed in the previous part.  As the velocity slices suggest,  this new zone is due to a relatively large separation bubble at the intersection of distal aortic arch and the descending aorta.  This unstable zone grows in size for the stressed model (cf.  TAVI0-Restored+S case) and produces stronger oscillations in the descending aorta (see figure \ref{fig:wss_restored}).  Whether such separation bubble emerges post AVR due to extreme out-of-plane bending of the aorta or dilation itself is the subject of a separate study.

\section{Conclusion}
We report a pattern separation of turbulent flow hemodynamics based on aortic stenosis patients with and without aortic dilation using GPU-accelerated image-based aortic flow simulations.  Three aortic stenosis cases,  one with aortic dilation and two within normal aortic size range were considered.  Vessel geometries were extracted from CT angiograms, and aortic inflow waveforms were obtained from echocardiography data.  All cases exhibited a turbulent flow with an elevated and oscillatory wall shear stress footprint on the ascending aorta.  However,  the dilated case showed a focal zone of elevated and oscillatory wall shear stress on the outer curvature of the aortic wall,  unlike other cases which exhibited a more extended distribution (i.e.,  covering the whole circumference of the vessel and further towards and even past the aortic arch).  Because the wall shear stress was elevated and oscillatory for both dilated and non-dilated cases,  it is suspected that another flow-mediated mechanism might be driving the dilation.  Further inspection showed that this mechanism could be a local pressure maximum adjacent to the elevated wall shear stress zone,  which is only present for the dilated case.  We argue that this persistent zone of elevated pressure is due to low velocity zone forming behind the aortic jet impingement site on the ascending aorta.  This observation is quantified using pressure probes,  where the non-dilated cases showed significantly stronger pressure oscillations without developing a focal zone of high pressure.  Moreover, we show that if this impingement zone is disturbed,  e.g.  via fluid suction from a more proximal brachiocephalic artery inlet,  it could lead to stronger turbulence with a more extended (less focal) elevated wall shear stress footprint.   This is examined using an idealized model of the aorta (a candy-cane geometry with uniform aortic radius),  where a more distal brachicephalic artery inlet was shown to lead to undisturbed jet impingement and thereby it developed focal zones of elevated wall shear-stress and pressure. Therefore,  it can be said that for the dilation case,  a focal zone of elevated pressure cooperates with a coinciding zone of elevated and oscillatory wall shear stress: the excessive and oscillatory wall shear stress weakens the arterial wall matrix,  while pressure buildup provides the driving normal force to dilate it.   

For the dilated case,  we further inspected the aortic hemodynamics after the aortic valve replacement using a virtually restored inflow model.  Simulation outcomes showed a laminar flow with no wall shear stress or pressure maxima in the ascending aorta, but instead revealed a zone of elevated wall shear stress at the distal aortic arch region.  Velocity slices at this area showed a separation bubble, which could be a result of dilation or out-of-plane remodelling of the aorta.  We showed that even an extreme case of stress (in which the peak velocity was doubled) does not lead to any focal area of elevated wall shear stress or pressure,  even though it led to elevated and oscillatory wall shear stress and pressure  in the ascending aorta. 

The distinct separation in wall hemodynamics between aortic stenosis patients with and without dilation that is presented here should be tested on a larger cohort of patients for further validation.  Moreover,  as we showed here,  a stronger turbulent flow may remove the stress from focal areas in the aorta,  thereby preventing the dilation.  However,  the key remaining question is, whether this prevention is a case of life-saving turbulence, or is it evil in disguise which can lead to sudden dissection through strong pressure oscillations? Lastly, it is important to establish a fluid mechanics criterion on occurrence of focal wall shear stress and pressure maxima (or jet impingement) in terms of peak velocity and jet radius (equivalent to effective orifice area).  We have seen that a larger jet,  even with peak velocities double than normal,  leads to no impingement on the wall, as it undergoes turbulent breakdown before arriving at the wall. This is contrary to the disease case, which involved a narrower jet with slightly higher peak velocity.

\label{secConc}

\section{Limitations}
The rigid wall assumption was used for all simulations presented here,  and is based on the correlation between present aortic atherosclerosis in the cases studied and aortic stiffness given in the literature.  However,  supporting kinematics data for the specific cases studied here was not available which could be a source of over-estimated wall pressure,  even though it is less likely to affect its distribution.  More accurate simulations can be performed by inferring wall kinematics from cine MRI scans.

Further,  the inflow jet profile was fabricated in this work based on a circular shape jet and fully stagnant peripheral part.  This could be improved by using PC MRI flow slices at the domain inflow,  which yields a better estimation of jet shape.

The pressure waveforms applied to the descending aorta outflow boundary were based on a simple two-element Windkessel model.  This could be improved by pressure data or more complex Windkessel models.

\section*{Acknowledgements}
HZ and RRK would like to thank the Swiss National Supercomputing Centre (CSCS) for providing technical support and GPU-node resources on the Cray XC40/50 supercomputer \textit{Piz Daint}. HZ would like to additionally acknowledge the financial support from Swiss National Science Foundation (SNSF) through the Early PostDoc Mobility Fellowship P2BEP2$\_191786$ and from the European Commission under the Marie Curie Individual Fellowship OPTAVI-895580.

\begin{figure}
\centering
\includegraphics[width=\textwidth]{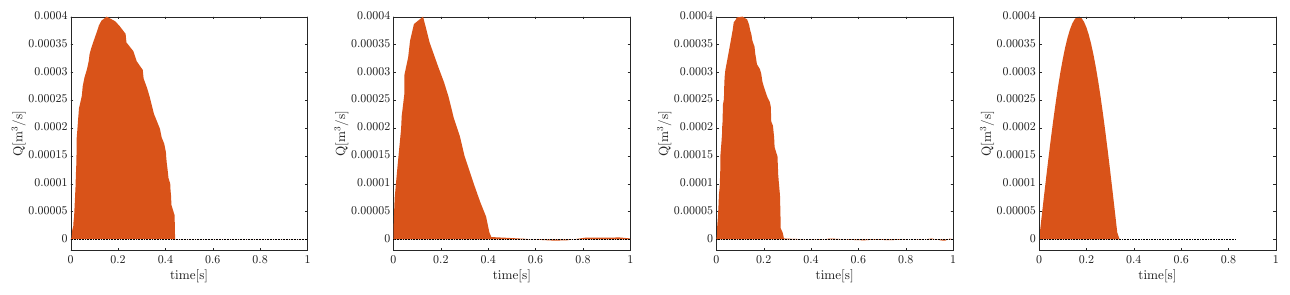}
%
%
\caption{Flow rate waveforms from left to right, for cases TAVI0, TAVI1, TAVI2 and Idealized.  }
\label{fig:flowrate_waveforms}
\end{figure}

\begin{figure}
\centering
\includegraphics[width=\textwidth]{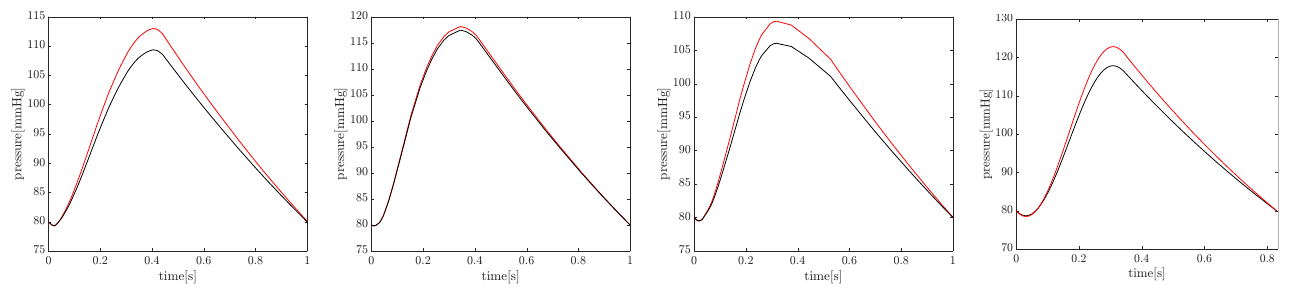}
%
%
\caption{Pressure waveforms, from left to right, for cases TAVI0, TAVI1, TAVI2 and Idealized.  Black curves show the waveform for the descending aorta outflow boundary, while the red curves show the corresponding waveforms at the inflow boundary.}
\label{fig:pressure_waveforms}
\end{figure}

 \appendix{\bf Appendix A: Flow rate and pressure waveforms} 
 \label{A}
 The velocity waveforms are taken from continuous wave (CW) spectral echocardiography data (see figure  \ref{fig:fig1}).  All physiological waveforms are projected based on a heartbeat of 60bpm (beats per minute).  For the idealized geometry case,  a heart rate of 72bpm is used to fabricate the velocity and pressure waveform.   All flow and pressure waveforms are shown in figures \ref{fig:flowrate_waveforms} and \ref{fig:pressure_waveforms} and,  while their corresponding velocity waveform can be obtained using the reference values given in Table\ref{tab:tabRefValues}.

 \begin{table}[htb]
    \caption{Flow rate and pressure waveform reference values}
    \label{tab:tabRefValues}
    \centering
        \begin{tabular}{c | S[table-format=2.2]|
                       *{3}{S[table-format=2.2]|}
                             S[table-format=2.2]
                              }
        \hline
        Case
                & {\makecell{$U_{peak}$\\ ($\si{m/s}$)}}
                    & {\makecell{$R_\textit{out}$ \\ ($\si{mmHg/cm3/s}$)}}
                        & {\makecell{ $C_\textit{out}$\\ ($\si{cm3/mmHg}$)}}
                        & {\makecell{$p_{systole}$\\ ($\si{mmHg}$)}}    
                             & {\makecell{$p_{diastole}$\\ $\si{mmHg}$}}                 \\  \hline
        TAVI 0 &  2.7  & 0.898 & 2.066 & 113 & 80  \\ \hline
        TAVI 1 & 2.1   & 1.210 & 1.351 & 118 & 80  \\ \hline
        TAVI 2 & 2.2  & 1.083 & 1.900 & 109 & 80  \\ \hline
        Idealized & 2.7  & 1.127 & 1.153 & 113 & 80  \\ \hline
           TAVI 0 Restored &  0.9  & 0.898 & 2.066 & 123 & 80  \\ \hline
                     TAVI 0 Restored+ &  1.8  & 0.6737 & 0.950 & 113 & 80  \\ \hline
        \end{tabular}
\end{table}

%
%


\bibliographystyle{vancouver}
\bibliography{mybibfile}

\renewcommand\thefigure{\arabic{figure}} 

\end{document}